%% file: main.tex
\documentclass[10pt,conference]{IEEEtran}
\usepackage[T1]{fontenc}
\usepackage{textcomp}
\usepackage{lmodern}
\usepackage{sty/main}
\usepackage{nicefrac}
\usepackage{sty/gnuplot-lua-tikz}

\title{\textbf{Optimal Software Architecture From Initial Requirements: An End-to-End Approach}}

\input{author}

\begin{document}

\maketitle

\begin{abstract}
\input{abstract}
\end{abstract}


\section{Introduction}\label{sec:introduction}
\input{introduction}

\section{Background}\label{sec:background}
\input{model}
\input{background}

\section{An End-to-end Method}\label{sec:method}
\input{method}

\section{Evaluation}\label{sec:evaluation}
\input{casestudy}

\input{results}

\subsection{Discussion and Threats to Validity}
\input{discussion}

\section{Related Work}
\input{related}

\section{Conclusion}
\input{conclusion}


\printbibliography 

\end{document}

%% file: tex/author.tex
\newcommand{\hl}[1]{#1}
\author{%
\IEEEauthorblockN{\hl{Ofir T. Erlich \quad David H. Lorenz}}
\IEEEauthorblockA{\hl{\textit{Dept. of Mathematics and Computer Science}}\\
\hl{\textit{The Open University of Israel}}\\
\hl{Ra'anana, Israel}\\
\hl{lorenz@openu.ac.il}}
}

%% file: tex/abstract.tex
A software architect turns system requirements into a suitable software architecture through an \theAOP.
However, how should
	the architect decide which quality improvement to prioritize,
	e.g., security or reliability?
In software product line,
    should a small improvement in multiple products
    be preferred over
    a large improvement in a single product?
Existing \theAOM{s} handle
	various steps in the process, but none of them systematically
	guides the architect in generating
	an optimal architecture from the initial requirements.
In this work we present 
    an end-to-end approach for generating
	an optimal software architecture for a single software product 
	and 
	an optimal family of architectures for a family of products.
We report on a case-study of applying 
    our approach to optimize 
    five industry-grade products in a real-life product line architecture, 
    where $359$ possible combinations of ten different quality efforts were prioritized.

%% file: tex/keywords.tex
Software architecture, product line, software architecture evaluation, {architecture optimization method}

%% file: tex/introduction.tex
\newcommand{\dfd}[1]{\DFD{#1}, \Fref{AOP}}
A software architecture is an important property of 
	a software system.
Choosing an \emph{optimal architecture}~(\OA) early
	in the software production lifecycle
	is likely to
	improve design quality, reduce overall costs, and help manage
	risks~\cite{Clements:2003:ESA}.
Settling on a suboptimal architecture, on the other hand, may result in 
	risks to the system (e.g., stability, performance, or security issues)
	and risks to the project (e.g., budget overrun).

A primary role of a {software architect} is to select an
	{optimal software architecture} that best meets the system's
	\emph{initial requirements}~(\IR)~\cite{Maier:2000:ASA}.
To fulfill this role the architect engages in an iterative
    \emph{\theAOP}~(AOP)~\cite{Clements:2003:ESA},
    depicted in \Fref{DFD}.
This process involves requirement analysis
	(both functional and extra-functional), taking into account the
	existing architecture (or lack of one), and  adhering to various
	constraints (such as budget limits).
Multiple alternative architectures are considered during the process,
	each with a different
	cost and a different level of compliance with the requirements.

The role of a \emph{software product line}~(\SPL)
    architect is even more complex. 
A \emph{product line architecture}~(\PLA) comprises  
    a core architecture,
    architectures of shared assets, 
    and particular architectures of specific products.
The {\AOP} effort in a \SPL{} must balance between
	efforts targeted at optimizing specific products
	and efforts targeted at improving all the products~\cite{dolan1998stakeholders}.
\input{fig_DFD}

\subsection{Architecture Optimization Process}

\Fref{AOP} depicts a broad overview of the
	flow of data in the {\AOP}.
The input to the process is the initial list of
	system requirements, denoted~{\IR}.
The process comprises three subprocesses:
	{architecture}
	\emph{evaluation}~(\Pref{1.0})
	\emph{generation}~(\Pref{2.0}),
	and \emph{selection}~(\Pref{3.0}),
	which can be further decomposed
	into subsubprocesses
	(henceforth, \emph{activities}\DEBUG{,~\Fref{AOE}}).
During the process additional
	data stores are produced
	(henceforth, \emph{byproducts}).
Eventually,
	an optimal software architecture,
	denoted~{\OA}, is generated as the
	output of the process.
\input{fig_AOP}

%



For each activity in the process,
	software architects have a plethora of
	\emph{architecture optimization methods}~(AOM{s})~\cite{Patidar:2015:SSA}
	at their disposal. 
An architect
   	can use, for example, \emph{formal concept analysis}~(FCA) 
	to generate an initial architecture from
	requirements~\cite{Haoues:2017:GSA}, or use \emph{scenario-based architecture 
	reengineering} (SBAR)~\cite{Bengtsson:1998:SSA}
	to generate alternative architectures, or use CMU's
	\emph{architecture tradeoff analysis method} (ATAM)~\cite{Kazman:2000:AMA}
	to evaluate the different
	alternatives and select the optimal architecture.

However,
	in spite of being a primary role of the software
	architect, previous studies do not
	cover the entire \AOP.
Half of the~30 \AOM{s} listed in~\Tref{BIG},\footnote{%
	The list of \AOM{s} in~\Tref{BIG} was compiled from
	several surveys that were conducted over the years:
		4 from 2004~\cite{babar2004comparison},
		7 from 2006~\cite{Mattsson:2006:SAE},
		8 from 2008~\cite{roy2008methods},
		2 from 2012~\cite{barney2012software,shanmugapriya2012software},
		7 from 2014~\cite{ameller2014assisting,gonzalez2014models},
		and~2 from 2020~\cite{Farshidi:2020:CSA,lytra2020quality}.}
	focus solely on architecture evaluation~(\Pref{1.0}),
	dealing only with activity~\ActivityII{}.
Only two~($\nicefrac{1}{15}$) handle initial
	architecture generation (activity~\ActivityI), and 
	only five~($\nicefrac{1}{6}$)
	handle alternative architecture generation (activity~\ActivityIII).
In all the rest~($\nicefrac{23}{30}$),
	the architect is
	left to generate the initial and alternative
	architectures~(\Pref{2.0}) with no supporting process.


A third of the \AOM{s} handle
	architecture selection (activity~\ActivityV).
In the rest~($\nicefrac{2}{3}$),
	the architect has to single out an optimal architecture~(\Pref{3.0})
	from a list of quality risks that those methods
	identify.
Some of the \AOM{s} do point out the need for
	one or more additional steps to complete the \AOP.
Yet, to our knowledge no method
	provides an end-to-end, systematic way to generate the optimal
	architecture from the initial requirements.
\input{tab_compare}

\subsection{Challenge}\label{sec:Challenge}
The architect cannot simply mix and match these \AOM{s}
    into a complete and impartial method for carrying out the \AOP.
    
First,
    many~($\nicefrac{7}{15}$) of the \AOM{s} are \emph{scenario-based}.
A common flaw of scenario-based methods is their reliance
    on the stakeholders' subjective opinion on technical issues~\cite{Ionita:2002:SSA}.
When a scenario-based model is used, 
    there is thus a risk that the {\AOP} will be biased by the
    predisposition of the organization decision-making process~\cite{Kazman:1999:EPA}.

Second, a majority~($\nicefrac{8}{15}$) of the \AOM{s} apply \emph{single-parameter optimization},
    namely, a single quality characteristic.
However, 
	in order to find an optimal architecture, a
	comparison of multiple quality characteristics
	is often required\DEBUG{~(\TraitVII)}.
For example, when a
	single-parameter \AOM{},
	such as the \emph{scenario based architecture level usability analysis}~(SALUTA)~\cite{folmer2004software},
	recommends an architecture due to it superior usability,
	the architect might not be able to use that recommendation
	in practice,
	as the architecture might have security, reliability, and
	performance issues, which the method does not 
	analyze.

Third, most~($\nicefrac{4}{5}$) of the \AOM{s} are \emph{cost-agnostic}.
When the cost of each architecture
	alternative is not taken into consideration\DEBUG{~(\TraitVIII)},
	the selection becomes
	biased towards the architecture with the highest
	quality improvement, even though it might be
	too expensive to implement or provide a too small
	a return on investment.

Fourth,
	the vast majority~($\nicefrac{13}{15}$) of the \AOM{s} on our list 
	assume a monolithic system
	and thus apply a \emph{single-objective optimization}.
However,
    in the case of a \SPL,    
    modifications to a \PLA{} can be made to either one of the 
    core, shared, or particular architectures~\cite{Faulk:2001:PRS}.
The architect needs to analyze these options simultaneously 
	in order to optimize the \PLA~\cite{Faulk:2001:PRS}.

Last,
	the various \AOM{s} are practically incompatible~\cite{Buchgeher:2010:ACM}.
They do not agree on the type of
	input/output, let alone on the quality model used.
For example,
	some \AOM{s} generate a textual output,
	while others assume as input a metric representation.
Some \AOM{s} do not define a quality model.
Others define their own unique model.
Therefore, the architect cannot easily combined them
	to create a full method
	for carrying out the \AOP.


\subsection{Contribution}
This work contributes a systematic, 
end-to-end approach to the \AOP.
In \Sref{method} we present a novel \AOM,
    named {\OAFIR} (\theOAFIR)
    that implements our approach.
{\OAFIR} includes all the activities
	\ActivityI,
	\ActivityII,
	\ActivityIII, and
	\ActivityV{}
	listed in \Tref{BIG},
	as well as all the properties
	\TraitVI,
	\TraitVII,
	\TraitVIII, and
	\TraitIX{}
	listed here:
\begin{itemize}
	\item\emph{Metrics-based}~(\trait~\TraitVI):
Suitability and quality gaps are rated in {\OAFIR}
	on a numerical scale.
The rating is done 
    with a separation of business decisions 
    from technical decisions.
Business stakeholders prioritize quality characteristics
    while the technical stakeholders calculate
    the impact of quality features 
    on each quality characteristic
    (like the impact of fault tolerance on reliability).
	
	\item\emph{Multi-parameter}~(\trait~\TraitVII):
{\OAFIR} uses a quality model 
    containing all extra-functional
    quality parameters from {ISO/IEC~25010}~\cite{ISO:25010:2011}, 
    simultaneously optimizing many quality features,
    while balancing, possibly conflicting, requirements.

	\item\emph{Cost-aware}~(\trait~\TraitVIII):
	The \AOP{} in {\OAFIR} takes into account
	the cost of each modification
	to the architecture,
	and optimizes the quality gain
	with a cost ratio or limit.
	
	\item\emph{Multi-objective}~(\trait~\TraitIX):
{\OAFIR} performs a metric comparison of the quality of different products in a \SPL,
    allowing the architect to 
    identify cross-product weaknesses,
    generate both product-specific and cross-product-line solutions,
    and reuse one product's effective solution across the \SPL.

\end{itemize}
To validate our approach, in \Sref{evaluation} we report  on a case study of 
applying {\OAFIR} to optimize a real-life, industry-grade \thePLA.

%% file: tikz/fig_DFD.tex
\begin{figure}
  \centering
  \input{tikz_DFD}
  \caption{The {\AOP}}
  \label{fig:DFD}
\end{figure}

%% file: tikz/tikz_DFD.tex
\pgfmathsetmacro{\myinnersep}{2}
\tikzset{%
	align=center,
	node distance=0.5cm and 0.5cm,
	ar/.style={->,>=latex},
	activity/.style={circle,
		text width=1.6cm,
		align=center,
		draw=black!80,
		fill=black!0,
		font=\scriptsize,
	},
	datastore/.style={
		text width=1.9cm,
		draw=black!60,
		very thick,
		shape=datastore,
		font=\scriptsize,
		inner sep=2,
		outer sep=0,
		align=center,
	},
	datastored/.style={trapezium,
		text width=1.9cm,
		align=center,
		draw=black!60,
		font=\scriptsize,
		inner sep=2,
		outer sep=0,
		shape=trapezium,
		trapezium left angle=70,
		trapezium right angle=110,
	},
}
\begin{tikzpicture}[%
]
\node[black,thick,dotted,fill=black!5,draw=black!60,text width=2cm,minimum height=1.5cm,minimum width=1.5cm,rectangle,pos=.5,font=\scriptsize] (dfd) {Architecture\\Optimization};
\node[datastore,left=of dfd,fill=yellow!5,font=\scriptsize,label=above:{\scriptsize\IR}] (ir) {Requirements};
\node[datastore,right=of dfd,fill=yellow!5,font=\scriptsize,label=above:{\scriptsize\OA}] (oa) {Optimal\\Architecture};
\draw[ar] (ir.south) to[bend right] (dfd.west);
\draw[ar] (dfd.east) to[bend left] (oa.north);
\end{tikzpicture}

%% file: tikz/fig_AOP.tex
\begin{figure}
  \centering
  \input{tikz_AOP}
  \caption{Subprocesses and the flow of data in the {\AOP}.}
  \label{fig:AOP}
\end{figure}
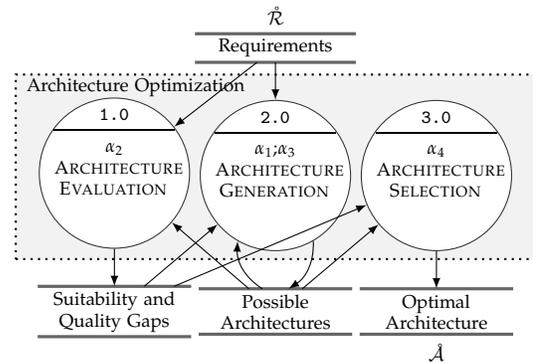

%% file: tikz/tikz_AOP.tex
\pgfdeclarelayer{bg}    
\pgfsetlayers{bg,main}  
%
\tikzset{%
	align=center,
	node distance=0.5cm and 0.1cm,
	ar/.style={->,>=latex},
	activity/.style={circle,
		text width=1.6cm,
		align=center,
		draw=black!80,
		fill=black!0,
		font=\scriptsize,
	},
	datastore/.style={
		text width=1.9cm,
		draw=black!60,
		very thick,
		shape=datastore,
		font=\scriptsize,
		inner sep=2,
		outer sep=0,
		align=center,
	},
	datastores/.style={trapezium,
		text width=1.9cm,
		align=center,
		draw=black!60,
		font=\scriptsize,
		inner sep=2,
		outer sep=0,
		shape=trapezium,
		trapezium left angle=70,
		trapezium right angle=110,
	},
}
\begin{tikzpicture}[%
]

\node[datastore,
] (b2) {Suitability and Quality Gaps};
\node[activity,above=of b2] (a2) {%
	\\[-2.5\baselineskip]\DFD{1.0}\\[-0.5\baselineskip]{\rule{\linewidth}{.4pt}}\\
	\BOX{\ActivityII}\\\sc Architecture Evaluation
};

\node[datastore,right=of b2,
] (b1) {Possible Architectures};
\node[activity,above=of b1] (a1) {%
	\\[-2.5\baselineskip]\DFD{2.0}\\[-0.5\baselineskip]{\rule{\linewidth}{.4pt}}\\
	\BOX{\ActivityI;\ActivityIII}\\\sc Architecture Generation
};

\node[datastore,right=of b1,fill=yellow!5,label=below:{\scriptsize\OA}] (oa) {Optimal Architecture};
\node[activity,above=of oa] (a5) {%
	\\[-2.5\baselineskip]\DFD{3.0}\\[-0.5\baselineskip]{\rule{\linewidth}{.4pt}}\\
	\BOX{\ActivityV}\\\sc Architecture Selection
};
\begin{pgfonlayer}{bg}    
\draw[black,thick,dotted,fill=black!5] ($(a2.north west)+(-0.6,0.6)$) rectangle ($(a5.south east)+(0.6,-0.4)$); 
\node (proc) at ([shift={(6.0:1.3)}]$(a2.north west)+(-0.3,0.3)$) [align=left] {\scriptsize Architecture~Optimization};
\end{pgfonlayer}

\node[datastore,above=of a1,minimum width=2.0cm,text width=2.0cm,fill=yellow!5,label=above:{\scriptsize\IR}] (ir) {Requirements}; 

\draw[ar,bend left,xshift=-0.5cm] (a1) edge (b1);
\draw[ar,bend left,xshift=0.5cm] (b1) edge (a1);
\draw[ar] (b1) edge (a2);
\draw[ar] (a2) -- (b2);
\draw[ar] (b2) -- (a1);
\draw[ar] (b2) -- (a5); 
\draw[ar] (a5) -- (oa);
\draw[ar] (ir) -- (a1);
\draw[ar] (ir) -- (a2);
\draw[ar] (b1) -- (a5); 
\end{tikzpicture}

%% file: tab/tab_compare.tex
\begin{table}
\definecolor{LightGray}{gray}{0.9}
\caption{Architecture optimization methods}
\label{table:BIG}
{\centering
\begin{tabular}
{|c@{}c|>{\centering}p{2em}@{}>{\centering}p{2em}@{}>{\centering}p{2em}@{}>{\centering}p{2em}|>{\centering}p{2em}@{}>{\centering}p{2em}@{}>{\centering}p{2em}@{}p{2em}|}
\hline
\multicolumn{2}{|c|}{}&
\multicolumn{4}{c|}{Activities}&
\multicolumn{4}{c|}{\Traits{}}
\tabularnewline
\hline 
\multicolumn{2}{|c|}{Method}&\ActivityI&\ActivityII&\ActivityIII&\ActivityV&\TraitVI&\TraitVII&\TraitVIII&\TraitIX\tabularnewline
\hline
\hline 
ABAS&\cite{klein1999attribute}&\NN&\YY&\NN&\NN&\NN&\YY&\NN&\NN\tabularnewline\rowcolor{LightGray}
AISAM&\cite{martens2010automatically}&\NN&\YY&\YY&\NN&\YY&\NN&\YY&\NN\tabularnewline
ALMA&\cite{lassing2002experiences}&\NN&\YY&\NN&\NN&\YY&\NN&\YY&\NN\tabularnewline\rowcolor{LightGray}
ARGUS-I&\cite{Vieira:2000:ASA}&\NN&\YY&\NN&\NN&\YY&\NN&\NN&\NN\tabularnewline
ASAAM&\cite{tekinerdogan2004asaam}&\NN&\YY&\NN&\NN&\NN&\NN&\NN&\NN\tabularnewline\rowcolor{LightGray}
ATAM&\cite{Kazman:2000:AMA}&\NN&\YY&\NN&\YY&\NN&\YY&\NN&\NN\tabularnewline
ATRIUM&\cite{montero2009atrium}&\YY&\NN&\NN&\NN&\NN&\YY&\NN&\NN\tabularnewline\rowcolor{LightGray}
CaLiPro&\cite{elorza2008evaluacion}&\NN&\YY&\NN&\YY&\NN&\YY&\NN&\YY\tabularnewline
CBAM&\cite{Kazman:2001:QCB}&\NN&\YY&\NN&\YY&\NN&\NN&\YY&\NN\tabularnewline\rowcolor{LightGray}
CBAMAHP&\cite{lee2009software}&\NN&\YY&\NN&\YY&\YY&\NN&\YY&\NN\tabularnewline
DoSAM&\cite{bergner2005dosam}&\NN&\YY&\NN&\YY&\YY&\YY&\YY&\NN\tabularnewline\rowcolor{LightGray}
D-SAAM&\cite{Graaf:2005:EES}&\NN&\YY&\NN&\NN&\NN&\YY&\NN&\YY\tabularnewline
EATAM&\cite{Kim:2008:EAA}&\NN&\YY&\NN&\YY&\NN&\YY&\NN&\YY\tabularnewline\rowcolor{LightGray}
EBAE&\cite{lindvall2003empirically}&\NN&\YY&\NN&\YY&\YY&\NN&\NN&\NN\tabularnewline
ESAAMI&\cite{molter1999integrating}&\NN&\YY&\NN&\NN&\NN&\NN&\NN&\NN\tabularnewline\rowcolor{LightGray}
FAAM&\cite{Dolan:2001:AAI}&\NN&\YY&\NN&\NN&\NN&\YY&\NN&\NN\tabularnewline
FCA&\cite{Haoues:2017:GSA}&\YY&\YY&\NN&\NN&\YY&\YY&\NN&\NN\tabularnewline\rowcolor{LightGray}
HoPLAA&\cite{olumofin2007holistic}&\NN&\YY&\NN&\YY&\NN&\YY&\NN&\YY\tabularnewline
LQN&\cite{petriu2000architecture}&\NN&\YY&\NN&\NN&\YY&\NN&\NN&\NN\tabularnewline\rowcolor{LightGray}
PASA&\cite{Williams:2002:PMP}&\NN&\YY&\YY&\YY&\NN&\NN&\YY&\NN\tabularnewline
QuaDAI&\cite{Gonzalez:2013:DVM}&\NN&\YY&\YY&\NN&\YY&\YY&\NN&\NN\tabularnewline\rowcolor{LightGray}
RARE&\cite{Barber:2002:EIS}&\NN&\YY&\NN&\NN&\YY&\NN&\NN&\NN\tabularnewline
SAAM&\cite{Kazman:1994:SMA}&\NN&\YY&\NN&\NN&\NN&\YY&\NN&\NN\tabularnewline\rowcolor{LightGray}
SAAMCS&\cite{lassing1999software}&\NN&\YY&\NN&\NN&\NN&\NN&\NN&\NN\tabularnewline
SAEM&\cite{duenas1998software}&\NN&\YY&\NN&\NN&\YY&\YY&\NN&\NN\tabularnewline\rowcolor{LightGray}
SALUTA&\cite{folmer2004software}&\NN&\YY&\NN&\NN&\YY&\NN&\NN&\NN\tabularnewline
SAM&\cite{wang1999introducing}&\NN&\YY&\NN&\NN&\YY&\NN&\NN&\NN\tabularnewline\rowcolor{LightGray}
SBAR&\cite{Bengtsson:1998:SSA}&\NN&\YY&\YY&\NN&\YY&\YY&\NN&\NN\tabularnewline
SPE&\cite{Williams:1998:PES}&\NN&\YY&\NN&\NN&\YY&\NN&\NN&\NN\tabularnewline\rowcolor{LightGray}
SQME&\cite{sedaghatbaf2019sqme}&\NN&\YY&\YY&\YY&\YY&\NN&\NN&\NN\tabularnewline
\hline 
\end{tabular}
\begin{tabular}{c l}
\ActivityI &
Initial architecture generation
\tabularnewline
\ActivityII &
Architecture evaluation
\tabularnewline
\ActivityIII &
Alternative architectures generation
\tabularnewline
\ActivityV &
Optimal architecture selection
\tabularnewline
\end{tabular}
\begin{tabular}{c l}
\hline 
\TraitVI &
Metrics-based assessment
\tabularnewline
\TraitVII &
Multi-parameter, multi-feature optimization
\tabularnewline
\TraitVIII &
Cost-aware optimization 
\tabularnewline
\TraitIX &
Multi-objective, multi-product optimization
\tabularnewline
\hline
\end{tabular}}
\end{table}

%% file: tex/model.tex
\newcommand{\Quality}[2]{\QUALITY\left({#1},{#2}\right)}
\newcommand{\SPLQuality}[3]{\QUALITY_{#1}\left({#2},{#3}\right)}
\newcommand{\QR}{{\mbox{\IR}}}
\renewcommand{\QR}{{\cal R}} 
\newcommand{\AR}{{\cal A}}
\newcommand{\MM}{{\cal M}}
\newcommand{\MS}{{{\cal M}^{\ast}}}
\newcommand{\MK}{{\MI_{K}}}
\newcommand{\Mi}{{\mu}}
\newcommand{\MI}{\mbox{\textsc{m}}}
\newcommand{\Cost}[1]{cost({#1})}
\newcommand{\TCR}{\kappa} 
\renewcommand{\TCR}{\xi} 

\subsection{Architecture Optimization}
Let~\(\Quality{\AR}{\QR}\in\R\) denote the overall quality adherence
	of an architecture \(\AR\in\DomainA\) to the system's quality
	requirements \(\QR\in\DomainR\),
	where \(\QUALITY\) is a real-valued function from~\(\DomainA\times\DomainR\) to~\(\R\) 
	and higher values mean better adherence.
Let~\(\MM=\{\Mi_i\in\DomainA\hookrightarrow\DomainA\}_{i\in I}\)
    denote a set of architecture modification options indexed by~$I$.

Given a cost estimation function,
    \(cost:\MM\rightarrow\R\),
    and a cost limit, \(\TCR\in\R\), 
	set by the stakeholders for architecture optimization,
    the goal
%
%
	of the software architect is to systematically
	find
	a subset of modifications
	\(
        \wp(\MM)\ni  
        \MK=
		\left\{\Mi_{i_k}\in\MM\right\}_{i_k\in K}
	\),
	$K\subset I$,
	that satisfies the cost constraint 
	\begin{equation}
	    \TCR\geq\sum_{{i_k}\in K}{\Cost{\Mi_{i_k}}}
	\label{equation:TCR}
	\end{equation}
	and maximizes 
	\(\Quality{\AR'}{\QR}\),
	where
	\[
	\AR'
	={\Mi_{i_k}\left(\cdots\Mi_{i_2}\left(\Mi_{i_1}\left(\AR\right)\right)\cdots\right)}
    \]
	is the architecture obtained
	by applying all the modifications 
	indexed by~$K$ to~\(\AR\).

For example,
for \(\MM = \left\{\Mi_1, \Mi_2, \Mi_3, \Mi_4\right\}\)
and the modification costs depicted in \Fref{TCR},
the set of modifications
\({\MI_{3,4}}=\left\{\Mi_3,\Mi_4\right\}\)
is optimal for \(\TCR=8\),
but
\({\MI_{2,3}}=\left\{\Mi_2,\Mi_3\right\}\)
is optimal for \(\TCR=6\).
\input{fig_TCR}

Alternatively, the stakeholders may decide not to set~\(\TCR\).
Instead, a parameter~\(\gamma\) is set,
	which represents the importance of \(\Quality{\AR}{\QR}\)
	to the stakeholders with respect to implementation costs.
A higher \(\gamma\) means willingness on their part to accept a 
	higher cost in return to a more significant quality improvement.
In such a case,
	the goal is to maximize:
\begin{equation}
	\frac{\left(\Quality{\AR'}{\QR}\right)^\gamma}{\sum_{{i_k}\in K}{\Cost{\Mi_{i_k}}}}
	\label{equation:gamma}
\end{equation}
where \(K\) is not empty.

In the example shown in \Fref{GAMSA}, for each \(\gamma\) the optimum is different:
\({\MI_2}\) is optimal for~\(\gamma=0.7\);
\({\MI_3}\) for~\(\gamma=1\);
\({\MI_{3,4}}\) for~\(\gamma=1.2\); and
\({\MI_{1,2,3,4}}\) is optimal for~\(\gamma=1.6\).

%% file: fig/fig_TCR.tex
\begin{figure}
  \centering
  \input{optimal_modifications}
  \caption{Example: optimal modifications per $\TCR$ and per $\gamma$}
  \label{fig:TCR} 
  \label{fig:GAMSA} 
\end{figure}
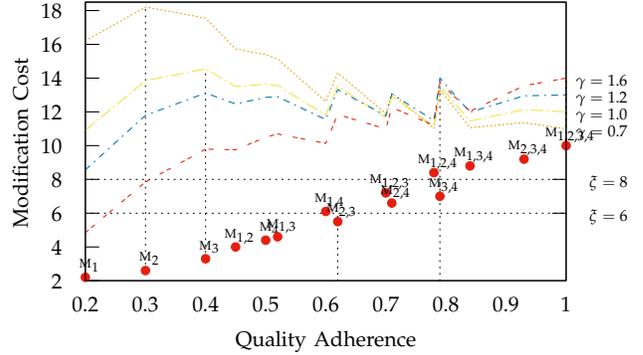

%% file: plot/optimal_modifications.tex
\begin{tikzpicture}[gnuplot]
\path (0.000,0.000) rectangle (9.000,5.000);
\gpcolor{color=gp lt color border}
\gpsetlinetype{gp lt border}
\gpsetdashtype{gp dt solid}
\gpsetlinewidth{1.00}
\draw[gp path] (1.136,0.985)--(1.316,0.985);
\node[gp node right,font={\fontsize{8.0pt}{9.6pt}\selectfont}] at (0.952,0.985) {$2$};
\draw[gp path] (1.136,1.434)--(1.316,1.434);
\node[gp node right,font={\fontsize{8.0pt}{9.6pt}\selectfont}] at (0.952,1.434) {$4$};
\draw[gp path] (1.136,1.883)--(1.316,1.883);
\node[gp node right,font={\fontsize{8.0pt}{9.6pt}\selectfont}] at (0.952,1.883) {$6$};
\draw[gp path] (1.136,2.333)--(1.316,2.333);
\node[gp node right,font={\fontsize{8.0pt}{9.6pt}\selectfont}] at (0.952,2.333) {$8$};
\draw[gp path] (1.136,2.782)--(1.316,2.782);
\node[gp node right,font={\fontsize{8.0pt}{9.6pt}\selectfont}] at (0.952,2.782) {$10$};
\draw[gp path] (1.136,3.231)--(1.316,3.231);
\node[gp node right,font={\fontsize{8.0pt}{9.6pt}\selectfont}] at (0.952,3.231) {$12$};
\draw[gp path] (1.136,3.680)--(1.316,3.680);
\node[gp node right,font={\fontsize{8.0pt}{9.6pt}\selectfont}] at (0.952,3.680) {$14$};
\draw[gp path] (1.136,4.129)--(1.316,4.129);
\node[gp node right,font={\fontsize{8.0pt}{9.6pt}\selectfont}] at (0.952,4.129) {$16$};
\draw[gp path] (1.136,4.579)--(1.316,4.579);
\node[gp node right,font={\fontsize{8.0pt}{9.6pt}\selectfont}] at (0.952,4.579) {$18$};
\node[gp node center,font={\fontsize{8.0pt}{9.6pt}\selectfont}] at (1.136,0.677) {$0.2$};
\node[gp node center,font={\fontsize{8.0pt}{9.6pt}\selectfont}] at (1.935,0.677) {$0.3$};
\node[gp node center,font={\fontsize{8.0pt}{9.6pt}\selectfont}] at (2.734,0.677) {$0.4$};
\node[gp node center,font={\fontsize{8.0pt}{9.6pt}\selectfont}] at (3.533,0.677) {$0.5$};
\node[gp node center,font={\fontsize{8.0pt}{9.6pt}\selectfont}] at (4.332,0.677) {$0.6$};
\node[gp node center,font={\fontsize{8.0pt}{9.6pt}\selectfont}] at (5.130,0.677) {$0.7$};
\node[gp node center,font={\fontsize{8.0pt}{9.6pt}\selectfont}] at (5.929,0.677) {$0.8$};
\node[gp node center,font={\fontsize{8.0pt}{9.6pt}\selectfont}] at (6.728,0.677) {$0.9$};
\node[gp node center,font={\fontsize{8.0pt}{9.6pt}\selectfont}] at (7.527,0.677) {$1$};
\draw[gp path] (7.527,1.883)--(7.347,1.883);
\node[gp node left,font={\fontsize{6.0pt}{7.2pt}\selectfont}] at (7.711,1.883) {${\TCR}=6$};
\draw[gp path] (7.527,2.333)--(7.347,2.333);
\node[gp node left,font={\fontsize{6.0pt}{7.2pt}\selectfont}] at (7.711,2.333) {${\TCR}=8$};
\draw[gp path] (1.136,4.691)--(1.136,0.985)--(7.527,0.985)--(7.527,4.691)--cycle;
\node[gp node left,font={\fontsize{6.0pt}{7.2pt}\selectfont}] at (7.527,3.006) {${\gamma=0.7}$};
\node[gp node left,font={\fontsize{6.0pt}{7.2pt}\selectfont}] at (7.527,3.231) {${\gamma=1.0}$};
\node[gp node left,font={\fontsize{6.0pt}{7.2pt}\selectfont}] at (7.527,3.456) {${\gamma=1.2}$};
\node[gp node left,font={\fontsize{6.0pt}{7.2pt}\selectfont}] at (7.527,3.680) {${\gamma=1.6}$};
\gpcolor{rgb color={0.000,0.000,0.000}}
\gpsetlinetype{gp lt axes}
\gpsetdashtype{gp dt axes}
\draw[gp path](1.136,1.883)--(7.527,1.883);
\draw[gp path](1.136,2.333)--(7.527,2.333);
\draw[gp path](4.491,0.985)--(4.491,1.883);
\draw[gp path](5.849,0.985)--(5.849,2.333);
\draw[gp path](1.935,1.120)--(1.935,4.627);
\draw[gp path](2.734,1.277)--(2.734,3.803);
\draw[gp path](5.849,2.108)--(5.849,3.679);
\draw[gp path](7.527,2.782)--(7.527,3.680);
\gpcolor{color=gp lt color border}
\node[gp node center,rotate=-270,font={\fontsize{8.0pt}{9.6pt}\selectfont}] at (0.292,2.838) {Modification Cost};
\node[gp node center,font={\fontsize{8.0pt}{9.6pt}\selectfont}] at (4.331,0.215) {Quality Adherence};
\node[gp node center,font={\fontsize{6.0pt}{7.2pt}\selectfont}] at (1.190,1.258) {$\MI_{1}$};
\gpcolor{rgb color={0.898,0.118,0.063}}
\gpsetpointsize{4.00}
\gppoint{gp mark 7}{(1.136,1.030)}
\gpcolor{color=gp lt color border}
\node[gp node center,font={\fontsize{6.0pt}{7.2pt}\selectfont}] at (1.989,1.348) {$\MI_{2}$};
\gpcolor{rgb color={0.898,0.118,0.063}}
\gppoint{gp mark 7}{(1.935,1.120)}
\gpcolor{color=gp lt color border}
\node[gp node center,font={\fontsize{6.0pt}{7.2pt}\selectfont}] at (2.788,1.505) {$\MI_{3}$};
\gpcolor{rgb color={0.898,0.118,0.063}}
\gppoint{gp mark 7}{(2.734,1.277)}
\gpcolor{color=gp lt color border}
\node[gp node center,font={\fontsize{6.0pt}{7.2pt}\selectfont}] at (3.187,1.662) {$\MI_{1,2}$};
\gpcolor{rgb color={0.898,0.118,0.063}}
\gppoint{gp mark 7}{(3.133,1.434)}
\gpcolor{color=gp lt color border}
\node[gp node center,font={\fontsize{6.0pt}{7.2pt}\selectfont}] at (3.587,1.752) {$\MI_{4}$};
\gpcolor{rgb color={0.898,0.118,0.063}}
\gppoint{gp mark 7}{(3.533,1.524)}
\gpcolor{color=gp lt color border}
\node[gp node center,font={\fontsize{6.0pt}{7.2pt}\selectfont}] at (3.746,1.797) {$\MI_{1,3}$};
\gpcolor{rgb color={0.898,0.118,0.063}}
\gppoint{gp mark 7}{(3.692,1.569)}
\gpcolor{color=gp lt color border}
\node[gp node center,font={\fontsize{6.0pt}{7.2pt}\selectfont}] at (4.386,2.134) {$\MI_{1,4}$};
\gpcolor{rgb color={0.898,0.118,0.063}}
\gppoint{gp mark 7}{(4.332,1.906)}
\gpcolor{color=gp lt color border}
\node[gp node center,font={\fontsize{6.0pt}{7.2pt}\selectfont}] at (4.545,1.999) {$\MI_{2,3}$};
\gpcolor{rgb color={0.898,0.118,0.063}}
\gppoint{gp mark 7}{(4.491,1.771)}
\gpcolor{color=gp lt color border}
\node[gp node center,font={\fontsize{6.0pt}{7.2pt}\selectfont}] at (5.184,2.381) {$\MI_{1,2,3}$};
\gpcolor{rgb color={0.898,0.118,0.063}}
\gppoint{gp mark 7}{(5.130,2.153)}
\gpcolor{color=gp lt color border}
\node[gp node center,font={\fontsize{6.0pt}{7.2pt}\selectfont}] at (5.264,2.246) {$\MI_{2,4}$};
\gpcolor{rgb color={0.898,0.118,0.063}}
\gppoint{gp mark 7}{(5.210,2.018)}
\gpcolor{color=gp lt color border}
\node[gp node center,font={\fontsize{6.0pt}{7.2pt}\selectfont}] at (5.823,2.650) {$\MI_{1,2,4}$};
\gpcolor{rgb color={0.898,0.118,0.063}}
\gppoint{gp mark 7}{(5.769,2.422)}
\gpcolor{color=gp lt color border}
\node[gp node center,font={\fontsize{6.0pt}{7.2pt}\selectfont}] at (5.903,2.336) {$\MI_{3,4}$};
\gpcolor{rgb color={0.898,0.118,0.063}}
\gppoint{gp mark 7}{(5.849,2.108)}
\gpcolor{color=gp lt color border}
\node[gp node center,font={\fontsize{6.0pt}{7.2pt}\selectfont}] at (6.303,2.740) {$\MI_{1,3,4}$};
\gpcolor{rgb color={0.898,0.118,0.063}}
\gppoint{gp mark 7}{(6.249,2.512)}
\gpcolor{color=gp lt color border}
\node[gp node center,font={\fontsize{6.0pt}{7.2pt}\selectfont}] at (7.022,2.830) {$\MI_{2,3,4}$};
\gpcolor{rgb color={0.898,0.118,0.063}}
\gppoint{gp mark 7}{(6.968,2.602)}
\gpcolor{color=gp lt color border}
\node[gp node center,font={\fontsize{6.0pt}{7.2pt}\selectfont}] at (7.581,3.010) {$\MI_{1,2,3,4}$};
\gpcolor{rgb color={0.898,0.118,0.063}}
\gppoint{gp mark 7}{(7.527,2.782)}
\gpcolor{rgb color={0.902,0.624,0.000}}
\gpsetlinetype{gp lt border}
\gpsetdashtype{gp dt 4}
\draw[gp path] (1.136,4.176)--(1.935,4.627)--(2.734,4.478)--(3.133,4.068)--(3.533,3.992)%
  --(3.692,3.934)--(4.332,3.368)--(4.491,3.750)--(5.130,3.209)--(5.210,3.481)--(5.769,3.008)%
  --(5.849,3.528)--(6.249,3.021)--(6.968,3.088)--(7.527,3.006);
\gpcolor{rgb color={0.941,0.894,0.259}}
\gpsetdashtype{gp dt 5}
\draw[gp path] (1.136,2.986)--(1.935,3.646)--(2.734,3.803)--(3.133,3.568)--(3.533,3.599)%
  --(3.692,3.583)--(4.332,3.187)--(4.491,3.574)--(5.130,3.156)--(5.210,3.435)--(5.769,3.039)%
  --(5.849,3.578)--(6.249,3.109)--(6.968,3.260)--(7.527,3.231);
\gpcolor{rgb color={0.000,0.447,0.698}}
\gpsetdashtype{gp dt 6}
\draw[gp path] (1.136,2.460)--(1.935,3.184)--(2.734,3.482)--(3.133,3.336)--(3.533,3.424)%
  --(3.692,3.432)--(4.332,3.129)--(4.491,3.527)--(5.130,3.179)--(5.210,3.469)--(5.769,3.116)%
  --(5.849,3.679)--(6.249,3.227)--(6.968,3.445)--(7.527,3.456);
\gpcolor{rgb color={0.898,0.118,0.063}}
\gpsetdashtype{gp dt 7}
\draw[gp path] (1.136,1.624)--(1.935,2.298)--(2.734,2.735)--(3.133,2.727)--(3.533,2.893)%
  --(3.692,2.937)--(4.332,2.812)--(4.491,3.197)--(5.130,3.004)--(5.210,3.290)--(5.769,3.051)%
  --(5.849,3.617)--(6.249,3.239)--(6.968,3.579)--(7.527,3.680);
\gpcolor{color=gp lt color border}
\gpsetdashtype{gp dt solid}
\draw[gp path] (1.136,4.691)--(1.136,0.985)--(7.527,0.985)--(7.527,4.691)--cycle;
\gpdefrectangularnode{gp plot 1}{\pgfpoint{1.136cm}{0.985cm}}{\pgfpoint{7.527cm}{4.691cm}}
\end{tikzpicture}

%% file: tex/background.tex
\input{fig_OFIR} 
\subsection{Optimization Methods}


An essential part of every \AOM{} is architecture 
	evaluation~(\Pref{1.0}).
An \AOM{} helps assessing the degree to which a
	software architecture fulfills the system's
	requirements, especially extra-functional
	requirements~\cite{Clements:2003:ESA}.
There are many kinds of \AOM{s}~\cite{roy2008methods},
	including:
    \emph{scenario-based},
    \emph{mathematical model-based},
    \emph{experience-based},
    \emph{simulation-based},
    \emph{metrics-based},
    \emph{tool-based}, and
    \emph{controlled experiments}.

While \AOM{s} vary in nature, they share a common structure~\cite{lee2009software}.
In every \AOM{}, some knowledge about the evaluated architecture(s) and the desired system's quality value(s) is processed, 
and the perceived suitability of the evaluated architecture(s) to the desired quality value(s) is produced as output.
Additional byproducts include prioritization of quality attributes, identification of architectural patterns used to compose the software architecture(s), and correlation between the architectural patterns in use to the quality attributes. 
These byproducts are intermediate results in some \AOM{s}.
Other \AOM{s} define them either as a expected input or as provided output.

While the majority of \AOM{s} are contented with just architecture evaluation, the \AOP{} is far from over for the architect. 
The software architect now needs to thoughtfully address the reported quality gaps of each architecture. 
Specifically, the architect must decide whether or not to:
send the recommended, most-suitable architecture to implementation;
apply some modifications to the architecture(s) to receive better results; or
take business actions in face of intolerable quality-risks.
Should the software architect decide to refine the architecture, a subset of the quality gaps is selected for minimization.


During architecture generation~(\Pref{2.0}),
the software architect identifies a list of changes to the former evaluated architecture that addresses some or all of the quality gaps. 
Since multiple quality gaps can be addressed in more than one way, multiple possible change lists exist.
For a single software system, the software architect can either
make a specific change to the architecture for each quality gap, or
make changes that address multiple quality gaps at once.

For example, consider the architecture of a system's web service that is diagnosed with quality gaps in four quality parameters: maturity, availability, modularity, and co-existence with external services. After analyzing the risks and possible solutions, the software architect has several options:
\begin{enumerate}
    \item Fixing each one of the 4 quality gaps separately, by adding metrics to the web service's availability, creating a stand-by service instance, separating the service discovery code from the business logic, and adding a throttling mechanism.
    \item Hosting the web service on public cloud virtual servers to rectify both maturity and availability risks, with the other quality risks handled as in item~1. This reduces the number of changes from 4 to 3 and has different consequences.
    \item Installing an API gateway in front of the web service to tackle all of the quality risks together.
\end{enumerate}
To proceed, another iteration of the \AOP{} is conducted in order to evaluate the suitability of the new architecture(s) in addressing selected quality gaps. In each iteration, the output of the \AOP{} is used to create an updated input for another iteration of the process, until the stakeholders decide to stop refining and proceed with the latest or best result.


Note that some changes might cause new quality
gaps to appear while solving others. Each change
list has its own cost, too. The cost is based on the
progress of implementation of the current architecture
(as changing an implemented system costs more than
changing architecture diagrams), and the effort needed
to implement the changes themselves. Different change
lists will leave the architecture with different lists
of quality gaps. The challenge is finding the list of
changes which maximizes the quality while minimizing
the implementation cost.
%

%% file: tikz/fig_OFIR.tex
\begin{figure*}[t!]
  \centering
  \input{tikz_OFIR}
  \caption{The \AOP{} in \OAFIR{}}
  \label{fig:OFIR}
\end{figure*}

%% file: tikz/tikz_OFIR.tex
\pgfdeclarelayer{bg}    
\pgfsetlayers{bg,main}  
%
\tikzset{%
	align=center,
	node distance=0.5cm and 0.1cm,
	ar/.style={->,>=latex},
	activity/.style={circle,
		text width=1.6cm,
		align=center,
		draw=black!80,
		fill=black!0,
		font=\scriptsize,
	},
	datastore/.style={
		text width=1.9cm,
		draw=black!60,
		very thick,
		shape=datastore,
		font=\scriptsize,
		inner sep=2,
		outer sep=0,
		align=center,
	},
	datastores/.style={trapezium,
		text width=1.9cm,
		align=center,
		draw=black!60,
		font=\scriptsize,
		inner sep=2,
		outer sep=0,
		shape=trapezium,
		trapezium left angle=70,
		trapezium right angle=110,
	},
}
\begin{tikzpicture}[%
]

\node[activity,label={[shift={(0.0,-1.7)}]\scriptsize\sref{STEP:I}}] (A1) {%
	\\[-2.5\baselineskip]\DFD{0.1}
	\\[-0.5\baselineskip]{\rule{\linewidth}{.4pt}}\\
	~
	\\\sc Quality Scope
};

\node[activity,right=of A1,label={[shift={(0.0,-1.7)}]\scriptsize\sref{STEP:II}}] (A2) {%
	\\[-2.5\baselineskip]\DFD{0.2}
	\\[-0.5\baselineskip]{\rule{\linewidth}{.4pt}}\\
	~
	\\\sc Quality\\Prioritiz.
};

\node[activity,right=of A2,label={[shift={(0.0,-1.7)}]\scriptsize\sref{STEP:III}}] (A3) {%
	\\[-2.5\baselineskip]\DFD{2.1}
	\\[-0.5\baselineskip]{\rule{\linewidth}{.4pt}}\\
	\BOX{\ActivityI}
	\\\sc Architecture Cons.
};

\node[activity,right=of A3,label={[shift={(0.0,-1.7)}]\scriptsize\sref{STEP:IV}}] (A4) {%
	\\[-2.5\baselineskip]\DFD{1.0}
	\\[-0.5\baselineskip]{\rule{\linewidth}{.4pt}}\\
	\BOX{\ActivityII}\\\sc Architecture Evaluation
};

\node[activity,right=of A4,label={[shift={(0.0,-1.7)}]\scriptsize\sref{STEP:V}}] (A5) {%
	\\[-2.5\baselineskip]\DFD{2.2}
	\\[-0.5\baselineskip]{\rule{\linewidth}{.4pt}}\\
	\BOX{\ActivityIII}\\\sc Architecture Generation
};

\node[activity,right=of A5,label={[shift={(0.0,-1.7)}]\scriptsize\sref{STEP:VI}}] (A6) {%
	\\[-2.5\baselineskip]\DFD{3.0}
	\\[-0.5\baselineskip]{\rule{\linewidth}{.4pt}}\\
	\BOX{\ActivityV}\\\sc Architecture Selection
};

\begin{pgfonlayer}{bg}    
\draw[black,thick,dotted,fill=black!5] ($(A1.north west)+(-0.6,0.6)$) rectangle ($(A6.south east)+(0.6,-0.4)$); 
\node (proc) at ([shift={(6.0:1.3)}]$(A1.north west)+(-0.3,0.3)$) [align=left] {\scriptsize Architecture~Optimization};
\end{pgfonlayer}

\node[datastore,above=of A1,fill=yellow!5,label=above:{\scriptsize}] (iso) {ISO/IEC 25010};
\node[datastore,above=of A2,minimum width=2.0cm,text width=2.0cm,fill=yellow!5,label=above:{\scriptsize\IR}] (ir) {Requirements}; 
\node[datastore,above=of A3,fill=yellow!5,label=above:{\scriptsize\Tref{products}}] (P) {Products};
\node[datastore,below=of A3,fill=yellow!5,label=above:{\scriptsize}] (IA) {Initial Architectures};

below Byproducts
\node[datastore,below=of A1,label=below:{\scriptsize\Tref{Quality:Properties}}] (B1) {Quality\\Model};
\node[datastore,below=of A2,label=below:{\scriptsize\Tref{Questions}}] (B2) {Compliance Questionnaire};
\node[datastore,below=of A4,fill=yellow!5,label=below:{\scriptsize\Tref{WResults}}] (B3) {Suitability and Quality Gaps};
	] (B4) {Suitability and Quality Gaps};
\node[datastore,below=of A5,fill=yellow!5,label=below:{\scriptsize\Tref{Modifications}}] (B5) {Alternative Architectures};
\node[datastore,below=of A6,fill=yellow!5,label=below:{\scriptsize\OA}] (B6) {Optimal Architecture};


\draw[ar] (iso) edge (A1);
\draw[ar] (ir) -- (A1);
\draw[ar] (ir) -- (A2);
\draw[ar] (A1) -- (B1);
\draw[ar] (A1) -- (B2);
\draw[ar] (B1) -- (A2);
\draw[ar] (A2) -- (B1) node[midway,sloped,below] {\tiny$\wc$};
\draw[ar] (B2) edge (A2);
\draw[ar] (B2) -- (A5);
\draw[ar] (A2) -- (B2) node[midway,sloped,below] {\tiny$\wf$};
\draw[ar] (B3) -- (A4);
\draw[ar] (A4) -- (B3) node[midway,sloped,below] {\tiny$Score$};
\draw[ar] (A4) -- (B3);
\draw[ar] (B3) -- (A5);
\draw[ar] (A5) -- (B5) node[midway,sloped,above] {\tiny$\MM$};
\draw[ar] (B5) -- (A6) node[midway,sloped,below] {\tiny$\MK$};
\draw[ar] (A6) -- (B6);
\draw[ar] (P) -- (A3);
\draw[ar] (A3) -- (IA);
\draw[ar] (IA) -- (A4);
\draw[ar] (IA) -- (A5);
\draw[ar] (B3) -- (A3); 
\draw[ar] (B2) -- (A4); 
\draw[ar] (P) -- (A1);
\draw[ar] (P) -- (A4); 
\draw[ar] (ir) -- (A3);

\end{tikzpicture}

%% file: tex/method.tex
In this section,
	we introduce our \AOM{} named \theOAFIR~(\OAFIR).
%
The method has 6 steps (\Fref{OFIR}):
\def\STEPI{Scope}
\def\STEPII{Prioritization}
\def\STEPIII{Construction}
\def\STEPIV{Evaluation}
\def\STEPV{Generation}
\def\STEPVI{Selection}
\emph{\STEPI}~(\Sref{STEP:I});
\emph{\STEPII}~(\Sref{STEP:II});
\emph{\STEPIII}~(\Sref{STEP:III});
\emph{\STEPIV}~(\Sref{STEP:IV});
\emph{\STEPV}~(\Sref{STEP:V}); and
\emph{\STEPVI}~(\Sref{STEP:VI}).

\subsection{\STEPI}\label{sec:STEP:I}
The first step in {\OAFIR} 
	sets the appropriate \emph{quality model} for the \AOP.
The architect 
	identifies quality characteristics
	that are relevant to all the software products
	being evaluated
	(\Sref{Characteristic:Selection}),
	and selects significant features 
	to be considered in their evaluation
	(\Sref{Feature:Selection}).

%
%

\input{tab_aspects}
\subsubsection{Characteristic Selection}\label{sec:Characteristic:Selection}

The architect starts with the quality model found in
	{ISO/IEC 25010}~\cite{ISO:25010:2011},
	which partitions the notion of software quality into eight quality characteristics:
%
%
	\iso{Functional suitability},
	\iso{Efficiency},
	\iso{Compatibility},
	\iso{Usability},
	\iso{Reliability},
	\iso{Security},
	\iso{Maintainability},
	\iso{Portability}.
Each characteristic is composed of several
	correlated quality properties.
To possess a
	quality characteristic, an architecture
	has to adhere to the quality properties it is
	composed of.

The ISO/IEC 25010 model is then narrowed down by
	eliminating characteristics irrelevant
	to software architecture 
	or irrelevant to the system.
For example, when the goal is 
	evaluation of both functional and extra-functional aspects of the system,
	all characteristics may be deemed relevant.
However, in the context of an {\AOP} the enterprise
	management and software architects may find the
	evaluation of extra-functional aspects to be more meaningful,
	in which case some of the characteristics may be eliminated from the quality model.

First, \iso{Functional suitability} can be ignored, because deciding on 
	an architecture is mostly the result of balancing trade-offs between extra-functional
	requirements.
Functional characteristics do not compete with
	extra-functional over which architecture to choose the same way different
	extra-functional characteristics do.

Second, \iso{Usability} can also be ignored.
Most of the properties of \iso{Usability} are functional,
	or otherwise not related to the software architecture.
The only two extra-functional properties of \iso{Usability}
	are \iso{User error-protection}, which is covered by the \iso{Fault
	tolerance} property of \iso{Reliability}, and \iso{Learnability} of
	code and structure, which is covered by the \iso{Analyzability} property of
	 \iso{Maintainability}. 

Third, \iso{Compatibility} 
	can be merged into the other characteristic.
We find its \iso{Interoperability} property inseparable from 
	\iso{Maintainability}, regarding consequences of changes in integrating
	components, and its \iso{Co-existence} property directly related to 
	\iso{Efficiency}. 
	
%
For the rest of this paper we shall assume
	that the five selected quality
	characteristics
	for extra-functional quality evaluation,
	and their associated properties,
	are those listed and numbered
	in~\Tref{Quality:Properties}.

\subsubsection{Feature Selection}\label{sec:Feature:Selection}


The properties, as defined in ISO/IEC 25010,
	can be evaluated in different ways.
In order to share the results among	products,
    which is the main goal of {\OAFIR},
	all the products of the \SPL{} 
	need to be evaluated with the same criteria and
	metrics~\cite{bergner2005dosam}.
To this end, key features
	which implement or compose each of the properties are identified 
	and defined in terms of the organization's common architectural patterns. 

For each such feature, the architect forms one or more yes-no
	survey questions. 
These questions should be articulated in such a way
	that answering \emph{yes} on all the questions related to a
	given property implies full compliance with 
	that property, while answering \emph{no} on all of them
	implies total neglect.

For example, the \iso{Fault tolerance} property of \iso{Reliability}
	may be broken down into five distinct features: protection from logical
	exceptions, wrong user actions, faults in
	external services, hardware failures, and full
	site failures;
in which case the following questions will
	be formed:
\begin{enumerate}
	\item[\scriptsize\ISO{1.3.1}] \Question{\iso{Fault tolerance (exc)}}{\small Does the product protect its activity from logical exceptions?}
	\item[\scriptsize\ISO{1.3.2}] \Question{\iso{Fault tolerance (user)}}{\small Does the product protect its activity from wrong user actions?}
	\item[\scriptsize\ISO{1.3.3}] \Question{\iso{Fault tolerance (ext)}}{\small Does the product protect its activity from faults in external services?}
	\item[\scriptsize\ISO{1.3.4}] \Question{\iso{Fault tolerance (hw)}}{\small Does the product protect its activity from hardware failures?}
	\item[\scriptsize\ISO{1.3.5}] \Question{\iso{Fault tolerance (site)}}{\small Does the product protect its activity from full site failures?}
\end{enumerate}

\input{tab_questions}
The collection of questions make up a \emph{quality compliance questionnaire} 
	(like the one presented in \Tref{Questions}). 
Not all properties may be relevant to every
	organization or to every  product --- only relevant
	properties should be measured.
However, when a characteristic is relevant to some but not all of the
	organization's products, it should nevertheless be included,
	and skipped when irrelevant.
We discuss
	several approaches to dealing with irrelevant
	characteristics in~\Sref{Weighting:Characteristics}.
In cases where the solution is mediocre
	or when the answer differs between several parts
	the evaluated architecture,
	the answers may also be somewhere in between \emph{yes} and \emph{no}.

\subsection{\STEPII}\label{sec:STEP:II}


The second step in {\OAFIR} is prioritization.
Let \mbox{$\FeatureSet=\{\aFeature_1,\aFeature_2,\ldots\,\aFeature_{n}\}$} 
be the set of features selected in~\Sref{STEP:I}, 
and let $\CharacteristicSet=\{\aCharacteristic_1,\aCharacteristic_2,\ldots,\aCharacteristic_l\}$
be the partition of~$\FeatureSet$ into characteristics.
Each characteristic~\mbox{$\aCharacteristic\in\CharacteristicSet$} is assigned a normalized weight~$\wc(\aCharacteristic)\in[0,1]$,
such that
\(
	1=\sum_{\aCharacteristic \in \CharacteristicSet}{\wc(\aCharacteristic)}
\),
and each feature $\aFeature\in\FeatureSet$ is assigned a normalized weight~$\wf(\aFeature)\in[0,1]$,
such that for all~$\aCharacteristic\in\CharacteristicSet$,
\(
	1=\sum_{\aFeature \in \aCharacteristic}{\wf(\aFeature)}
\).
Giving due weighting to quality characteristics in~$\CharacteristicSet$ is 	a business decision, 
    while weighting features in~$\FeatureSet$ is an engineering decision.

\subsubsection{Weighting Characteristics}\label{sec:Weighting:Characteristics}
By assigning weights to the different characteristics,
	the organization management expresses which
	quality characteristic is of higher business priority,
	e.g., the management may choose to
	prioritize \emph{security} over \emph{performance}
	\cite{svahnberg2005investigation}.
Through updates to these priorities
	management can direct
	the organization towards the desired
	results \cite{ALNAEEM:2005:QSA}. 

By default,
	every characteristics~$\aCharacteristic\in\CharacteristicSet$ is assigned the 
	weight~$\nicefrac{1}{|\CharacteristicSet|}$
		($\nicefrac{1}{5}$ in~\Tref{Quality:Properties}).
An \AOP{} based on these default weights produces
	balanced quality scores for the various products.
Later, these weights can be changed according
	to business prioritization,
	and the revised evaluation scores would indicate which
	architectural changes 
	can
	make the biggest impact. 

\subsubsection{Weighting Features}\label{sec:Weighting:Features}
Feature weighting, on the other hand,
	is of a more technical nature,
	prioritizing elements of a desired
	product architecture.
These weights are much less likely to change
    during the \AOP{},
	unless a major cross-product technological
	change occurs, because they are 
	based more on advances in the state-of-the-art
	than on human decisions.

The weight of each feature 
	reflects the quality gained to the relevant characteristic by implementing the
	feature (or lost by not implementing it).
In weighting the different features the architect should consult
	professional studies and veteran architects.
For all $\aCharacteristic\in\CharacteristicSet$,
for all $\aFeature\in\aCharacteristic\subset\FeatureSet$,
we define the overall weight of feature~$\aFeature$
to be:
\begin{equation}
	w(\aFeature) \triangleq 100*\wc(\aFeature)*\wf(\aFeature)
\end{equation}
where 
$\wc(\aFeature)\equiv\wc(\aCharacteristic)$
is the weight assigned to the characteristic~$\aCharacteristic$ to which~$\aFeature$ belongs.

\subsection{\STEPIII}\label{sec:STEP:III}
To begin the \AOP{},
	an initial architecture is needed.
In the case of a \SPL,
	several products are chosen to present their initial architectures.
Note that selecting a subset of a \SPL's
	products can be sufficient for identifying
	common quality flaws across the product line~\cite{elorza2008evaluacion}.

If an initial architecture does not exist,
	one has to be generated.
A naive approach is
	to use appropriate design patterns for all the requirements,
	based on the quality compliance questionnaire
	from \Sref{Feature:Selection}.
This could produce a perfect or near-perfect architecture,
	but not necessarily a feasible one.
Applying all desired modifications
	might not adhere to the budget constraints.
It will be hard to backstep
	from the perfect architecture
	to an optimal feasible architecture.

Instead,
	in {\OAFIR}
	the most basic design patterns
	are applied to create a minimal initial
	architecture, with only those patterns that are necessary
	to fulfill the functional requirements.
The quality gaps in this initial architecture
	will be identified downstream,
	and a set of modifications
	will be applied
	to achieve the optimal architecture.

\subsection{\STEPIV}\label{sec:STEP:IV}
In the fourth step of {\OAFIR}, 
	the specific architecture of each product is assessed by asking its architect in an interview
	the questions found in the quality compliance questionnaire (\Sref{Feature:Selection}).
Let
$\ProductSet=\left\{\aProduct_1,\aProduct_2,\ldots,\aProduct_m\right\}$
be a set of $m$ products of the \SPL{} being evaluated.
During the interviews, every feature~$\aFeature \in \FeatureSet$
in every product~$\aProduct \in \ProductSet$
is assigned a numeric value 
$\questValue{\aFeature}{\aProduct}\in{\Q}\cap[0,1]$ that reflects
$\aProduct$'s degree of compliance with~$\aFeature$.
	
When a certain feature~$\aFeature$ is irrelevant
	with respect to a specific product~$\aProduct$, 
	its weight~$\wf(\aFeature)$ can either be divided among the other features in the same characteristic
	or retained. 
If retained, $\questValue{\aFeature}{\aProduct}$ can be assigned either a perfect value, $\questValue{\aFeature}{\aProduct}=1$,
		indicating that there is no flaw in the architecture with respect to this feature, or
		an empty value, $\questValue{\aFeature}{\aProduct}=0$,
		indicating that the architecture does not handle that feature.
The last option is usually for a feature
		for which it is inherently difficult to find an architectural solution.
This way, products that are not concerned with that feature
		will not standout as those who do handle it successfully.

For all $\aFeature \in \FeatureSet$, $\aProduct\in\ProductSet$,
the weighted score for feature~$\aFeature$ in product~$\aProduct$ is defined to be:
\begin{equation}
	\weightedVal\triangleq w(\aFeature)*\questValue{\aFeature}{\aProduct}
\end{equation}
To detect exceptional scores,
for each feature~$\aFeature \in \FeatureSet$
we compute the mean score
\(\mean{\aFeature}=\frac{1}{|\ProductSet|}\sum_{\aProduct\in \ProductSet}{\weightedVal}\)
across products,
and its standard deviation,
$\stddev{\aFeature}$.
%
The quality gap of~$\aFeature$ is defined to be:
 \(\Delta_{\aFeature} \triangleq w(\aFeature) - \mean{\aFeature}\).
Features with a quality gap
\(\Delta_{\aFeature} > \mean{\Delta_{\aFeature}} + \stddev{\Delta_{\aFeature}}\)
have the highest 
impact over the quality of the product line.
Features with \mbox{\(\weightedVal < \mean{\aFeature} - \stddev{\aFeature}\)}
have a major quality gap in a specific product~$\aProduct$.

\subsection{\STEPV}\label{sec:STEP:V}
The fifth step in {\OAFIR}
	is to generate the set of \emph{architecture modification options}, $\MM$~(recall \Sref{background}),
	that if applied to the existing architecture may address the quality gaps.

\subsubsection{Product Line Level Modifications}\label{sec:ProductLineMods}
%
For each high-impact feature,
one or more modifications are generated,
using as guide the questions formed in \Sref{Feature:Selection}.
In principle these modifications should achieve full adherence
to those quality features, i.e.,
$\forall{\aProduct}\in \ProductSet, \questValue{\aFeature}{\aProduct} = 1$.
In practice, however, sometimes a modification
only achieves a partial improvement.
Moreover, a modification can have a positive or negative impact
on other quality features~\cite{Niu:2009:CAP}.

There are different types of modifications in a \SPL{}.
%
%
Each modification incurs a total cost, which may
include a shared cost for the \SPL{} level effort
and particular costs for each product level effort.
Similarly, each modification also has its own total
quality gain, which includes the sum of quality
improvements in all quality feature in all of the
products.

\subsubsection{Product Level Modifications}\label{sec:SingleProdMods}
%
For each feature with a major quality gap,
one or more modifications are generated,
using the questions formed in \Sref{Feature:Selection}.
As opposed to \Sref{ProductLineMods},
only product-specific modifications
are analyzed at this step.
Still,
a modification may have an impact
on more than one quality feature.

Finally, each subset of these modifications
is assigned its total quality adherence gain
and total cost.
The full list of modifications is used
in the final step of {\OAFIR}.

\subsection{\STEPVI}\label{sec:STEP:VI}
The final step of {\OAFIR} 
provides the software architect
with a subset of modifications, $\MK\subset\MM$,
that produce the optimal architecture,
where $\MM$ is the list of modifications
derived in \Sref{STEP:V}.

In principle, there are~$2^{|\MM|}$ subsets in~$\wp(\MM)$ to consider.
In practice, the number is much lower because some of the modifications may be mutually exclusive. 
The subset with the highest overall quality
that adheres to the total cost restriction~$\TCR$~(\EQref{TCR}, \Sref{background})
is selected.
Alternatively, if~$\TCR$ is not set, the subset which maximizes the~$\gamma$ function
(\EQref{gamma}, \Sref{background}) is selected.
Applying the selected subset of modifications to
the existing architecture produces the optimal
architecture.

%% file: tab/tab_aspects.tex
\begin{table*}
\begin{center}
\caption{Quality characteristics selected from ISO/IEC 25010 categorization of software quality requirements}
\label{table:Quality:Properties}
\centering\scriptsize
\begin{tabular}{r|l|l|l|l|l| } 
\cline{2-6}
&\multicolumn{5}{c|}{\textbf{Software Product Quality Characteristics}}\\

&
\multicolumn{1}{c}{$\wc=\nicefrac{1}{5}$} & 
\multicolumn{1}{c}{$\wc=\nicefrac{1}{5}$} & 
\multicolumn{1}{c}{$\wc=\nicefrac{1}{5}$} & 
\multicolumn{1}{c}{$\wc=\nicefrac{1}{5}$} & 
\multicolumn{1}{c|}{$\wc=\nicefrac{1}{5}$} \\ 

\cline{2-6}
&
1. \ISO{Reliability} & 
2. \ISO{Efficiency} & 
3. \ISO{Maintainability} & 
4. \ISO{Portability} & 
5. \ISO{Security} \\

\cline{2-6}
\multirow{6}{*}{} &
1.1 \ISO{Maturity} &	
2.1 \ISO{Time behavior} &	
3.1 \ISO{Modularity} &	
4.1 \ISO{Adaptability} &	
5.1 \ISO{Confidentiality} \\

&
1.2 \ISO{Availability} &	
2.2 \ISO{Resource utilization} &	
3.2 \ISO{Reusability} &	
4.2 \ISO{Installability} &	
5.2 \ISO{Integrity} \\

&
1.3 \ISO{Fault tolerance} &	
2.3 \ISO{Capacity} &	
3.3 \ISO{Analyzability} &	
4.3 \ISO{Replaceability} &	
5.3 \ISO{Non-repudiation} \\

&
1.4 \ISO{Recoverability} &	
2.4 \emph{\ISO{Co-existence}}\footnotemark\addtocounter{footnote}{-1} &
3.4 \ISO{Modifiability} &
&	
5.4 \ISO{Accountability} \\

&
& &	
3.5 \ISO{Testability} &
& 
5.4 \ISO{Authenticity} \\

&
& &	
3.6 \emph{\ISO{Interoperability}}\tablefootnote{%
We consider \ISO{Coexistence} and \ISO{Interoperability} as properties of \ISO{Performance Efficiency} and \ISO{Maintainability}, respectively, although ISO/IEC 25010 places them in a separate 
category named \ISO{Compatibility}.} &	
& \\

\cline{2-6}
\end{tabular}
\end{center}
\end{table*}

%% file: tab/tab_questions.tex
\newcommand*\pied[1]{\tikz[
	pie values/.style={font={\tiny}},
	scale=.015
]{%
 \pie [%
 	rotate=180,%
	color={black!10},%
	text=pin,%
	before number=,
	radius =50]
	{#1}}
}
\renewcommand*\pied[1]{}

\begin{table*}
\begin{center}
\caption{Quality Compliance Questionnaire}
\label{table:Questions}
\centering
\begin{tabular}
{|@{}>{\bf\tiny}c@{}|@{}>{\scriptsize}l@{}>{\bf\scriptsize}r@{}|>{\sl}l@{}|c|}

\hline
\multicolumn{3}{|c|}{Feature}&Question&Weight\\
\hline
\multirow{8}{*}{\begin{sideways}Reliability\end{sideways}}&
        1.1 & {Maturity} & {Is the reliability of the product monitored? } 
& .14 
\\
        &1.2 & {Availability} & {Can the product be upgraded without downtime?} & .10 \\
        &1.3.1 & {Fault tolerance (exc)} & {Does the product protect its activity from logical exceptions?} & .14 \\
        &1.3.2 & {Fault tolerance (user)} & {Does the product protect its activity from wrong user actions?} & .14 \\
        &1.3.3 & {Fault tolerance (ext)} & {Does the product protect its activity from faults in external services?} & .14 \\
        &1.3.4 & {Fault tolerance (hw)} & {Does the product protect its activity from hardware failures?} & .14 \\
        &1.3.5 & {Fault tolerance (site)} & {Does the product protect its activity from full site failures?} & .06 \\
        &1.4 & {Recoverability} & {Does the product have a verified data backup and recovery ability?} & .14 \\

\hline
\multirow{6}{*}{\begin{sideways}Efficiency\end{sideways}}&
        2.1.1 & {Time behavior (req)} & {Does the product meet its latency and throughput requirements?} & .15 \\
        &2.1.2 & {Time behavior (mon)} & {Are the latency and throughput of the product's activity monitored?} & .15 \\
        &2.2.1 & {Resource utilization (req)} & {Does the product meet its resource utilization requirements?} & .15 \\
        &2.2.2 & {Resource utilization (mon)} & {Is the resource utilization of the product's activity monitored?} & .15 \\
        &2.3 & {Capacity} & {Is the product's activity load actively shared amongst its resources?} & .20 \\
        &2.4 & {Co-existence} & {Does the product prevent external load from harming its performance?} & .20 \\
 
\hline
\multirow{6}{*}{\begin{sideways}Maintainability\end{sideways}}&
        3.1 & {Modularity} & {Does a change impact only one component of the product?} & .20 \\
        &3.2 & {Reusability} & {Are external data and logic re-used (rather than re-implemented)?} & .14 \\
        &3.3 & {Analyzability} & {Is the architecture kept clear by adhering to common patterns?} & .14 \\
        &3.4 & {Modifiability} & {Do all product's services implement forward/backward-compatibility?} & .14 \\
        &3.5 & {Testability} & {Is the product continuously tested to supply confidence in changes?} & .19 \\
        &3.6 & {Interoperability} & {Are the inter-communication protocols loosely-coupled?} & .19 \\
 
\hline
\multirow{4}{*}{\begin{sideways}Portability\end{sideways}}&
        4.1 & {Adaptability} & {Can the product be scaled for higher usage by merely adding resources?} & .28 \\
        &4.2 & {Installability} & {Can the product be installed in different environments?} & .28 \\
        &4.3.1 & {Replaceability (lang)} & {Can the product's components be developed in different programming languages?} & .16 \\
        &4.3.2 & {Replaceability (inst)} & {Can the product be used without installing specific technologies?} & .28 \\
 
\hline
\multirow{5}{*}{\begin{sideways}Security\end{sideways}}&
        5.1 & {Confidentiality} & {Does the product use common authorization mechanisms?} & .15 \\
        &5.2 & {Integrity} & {Do the product's components prevent unauthorized access?} & .25 \\
        &5.3 & {Non-repudiation} & {Does the product permanently audit all user actions?} & .25 \\
        &5.4 & {Accountability} & {Does the product enforces use of personal users?} & .15 \\
        &5.5 & {Authenticity} & {Does the product fully-separate simulated data from real data?} & .20 \\

\hline
\end{tabular} 
\end{center}
\end{table*} 

%% file: tex/casestudy.tex
\input{tab_results}
{\OAFIR} was applied to an industrial \theSPL{} comprising dozens of products.
Five products,
	code-named \ProdA, \ProdB, \ProdC, \ProdD, and \ProdE,
    were randomly chosen for the {\AOP}. 
These five products
	had different requirements, constraints, and
	architectures, 
	and varied in their ``nature"~(\Tref{products}),
	e.g, in the number of users (ranging from few to thousands),
	in the rate of transactions (ranging from hundreds per hour to thousands a second),
	and in the number of interoperable systems (ranging from several to dozens of interfaces).
\input{tab_product}

%
Three expert architects were consulted 
	for the selection of features,
	assigning weights 
	representing the normalized impact that each feature has over its quality characteristic
	(left-hand section in \Tref{FWeights}).
The resulted quality compliance questionnaire is the one presented in \Tref{Questions}.
For each of the quality features,
	they also defined
	one or more architectural patterns
	that can be used to achieve the desired quality.

%% file: tab/tab_results.tex
\newcommand{\Std}{\mbox{\scriptsize$\mu\!\pm\!\sigma$}}
\newcommand{\Gap}{\scriptsize$\Delta$}

\begin{table*}
\begin{center}
\definecolor{LightGray}{gray}{.9}
\caption{Quality Scores}
\label{table:FWeights} 
\label{table:QResults} 
\label{table:WResults} 
{\centering\footnotesize
\begin{tabular}
{|@{}>{\bf\tiny}c@{}|@{}>{\scriptsize}l@{}>{\bf\scriptsize}r@{}|c|
|
>{\columncolor{colorA}\color{white}}c
>{\columncolor{colorB}}c
>{\columncolor{colorC}}c
>{\columncolor{colorD}}c
>{\columncolor{colorE}}c
|
|*{5}{c}|rlc|}
\hline
\rowcolor{LightGray}
\multicolumn{4}{|c||}{Assigned Weights}&
\multicolumn{5}{ c||}{Compliance}&
\multicolumn{8}{ c|}{Weighted Score} \tabularnewline
\hline
\rowcolor{LightGray}
\multicolumn{3}{|l}{Feature} & $\wf$ 
 & \ProdA & \ProdB & \ProdC & \ProdD & \ProdE 
 & \ProdA & \ProdB & \ProdC & \ProdD & \ProdE & \multicolumn{2}{c}{\Std} & \Gap\tabularnewline 

\hline
\hline
\multirow{8}{*}{\begin{sideways}Reliability\end{sideways}}&
1.1 & Maturity & .14
 & 1 & \textthreequarters & \textonehalf & \textthreequarters & 1
 & \mygreen{2.8} & 2.1 & \myred{1.4} & 2.1 & \mygreen{2.8} & 1.71 & 2.76 & \mygreen{.56}\tabularnewline
&
1.2 & Availability & .10
 & 1 & \textonehalf & \textthreequarters & 0 & \textonequarter 
 & \mygreen{2} & 1 & 1.5 & \myred{0} & .5 & .29 & 1.7 & 1\tabularnewline
&
1.3.1 & Fault~tolerance~(exc) & .14
 & 1 & 0 & 1 & 1 & 1
 & 2.8 & \myred{0} & 2.8 & 2.8 & 2.8 & 1.12 & 3.36 & \mygreen{.56}\tabularnewline
&
1.3.2 & Fault~tolerance~(user) & .14
 & 0 & 1 & 0 & 1 & 0
 & 0 & \mygreen{2.8} & 0 & \mygreen{2.8} & 0 & -.25 & 2.49 & 1.68\tabularnewline
&
1.3.3 & Fault~tolerance~(ext) & .14
 & 1 & 0 & 1 & 1 & 1
 & 2.8 & \myred{0} & 2.8 & 2.8 & 2.8 & 1.12 & 3.36 & \mygreen{.56}\tabularnewline
&
1.3.4 & Fault~tolerance~(hw) & .14
 & 1 & 1 & 0 & 1 & 1
 & 2.8 & 2.8 & \myred{0} & 2.8 & 2.8 & 1.12 & 3.36 & \mygreen{.56}\tabularnewline
&
1.3.5 & Fault~tolerance~(site) & .06
 & \textonethird & \texttwothirds & 0 & \textonethird & \texttwothirds 
 & .4 & \mygreen{.8} & \myred{0} & .4 & \mygreen{.8} & .17 & .78 & .72\tabularnewline
&
1.4 & Recoverability & .14
 & \texttwothirds & 1 & 0 & 1 & 1
 & 1.9 & 2.8 & \myred{0} & 2.8 & 2.8 & .96 & 3.14 & .74\tabularnewline
\hline
\rowcolor{LightGray}
\multicolumn{4}{>{\bf\tiny}r}{1.00}&
\multicolumn{5}{>{\bf\tiny}r}{
Reliability} & \multicolumn{1}{>{\bf\tiny}c}{15.5} & \multicolumn{1}{>{\bf\tiny}c}{12.3} & \multicolumn{1}{>{\bf\scriptsize}c}{\textcolor{myred}{8.5}} & \multicolumn{1}{>{\bf\tiny}c}{16.5} & \multicolumn{1}{>{\bf\tiny}c}{15.3} &&& \multicolumn{1}{>{\bf\scriptsize}c}{\textcolor{mygreen}{6.38}} \tabularnewline 
\hline
\multirow{6}{*}{\begin{sideways}Efficiency\end{sideways}}&
2.1.1 & Time~behavior~(req) & .15
 & 1 & 1 & 0 & \textthreequarters & 1
 & 3 & 3 & \myred{0} & 2.2 & 3 & 1.08 & 3.41 & .75\tabularnewline
&
2.1.2 & Time~behavior~(mon) & .15
 & 0 & 0 & 0 & 0 & 0
 & 0 & 0 & 0 & 0 & 0 & 0 & 0 & \myred{3}\tabularnewline
&
2.2.1 & Resource~utilization~(req) & .15
 & 1 & \textthreequarters & \textthreequarters & \textthreequarters & 1
 & \mygreen{3} & 2.2 & 2.2 & 2.2 & \mygreen{3} & 2.18 & 2.91 & \mygreen{.45}\tabularnewline
&
2.2.2 & Resource~utilization~(mon) & .15
 & 0 & 0 & 0 & 0 & 0
 & 0 & 0 & 0 & 0 & 0 & 0 & 0 & \myred{3}\tabularnewline
&
2.3 & Capacity & .20
 & \textthreequarters & 1 & \textonehalf & 1 & \textonehalf 
 & 3 & \mygreen{4} & \myred{2} & \mygreen{4} & \myred{2} & 2.1 & 3.89 & 1\tabularnewline
&
2.4 & Co-existence & .20
 & 1 & 0 & 1 & 1 & \textonehalf 
 & 4 & \myred{0} & 4 & 4 & 2 & 1.2 & 4.4 & 1.2\tabularnewline
\hline
\rowcolor{LightGray}
\multicolumn{4}{>{\bf\tiny}r}{1.00}&
\multicolumn{5}{>{\bf\tiny}r}{Efficiency} & \multicolumn{1}{>{\bf\scriptsize}c}{\textcolor{mygreen}{13}} & \multicolumn{1}{>{\bf\tiny}c}{9.2} & \multicolumn{1}{>{\bf\scriptsize}c}{\textcolor{myred}{8.2}} & \multicolumn{1}{>{\bf\scriptsize}c}{\textcolor{mygreen}{12.5}} & \multicolumn{1}{>{\bf\tiny}c}{10} &&& \multicolumn{1}{>{\bf\tiny}c}{9.4} \tabularnewline 
\hline
\multirow{6}{*}{\begin{sideways}Maintainability\end{sideways}}&
3.1 & Modularity & .20
 & 0 & 0 & 0 & 1 & \textonehalf
 & 0 & 0 & 0 & \mygreen{4} & 2 & -.4 & 2.8 & \myred{2.8}\tabularnewline
&
3.2 & Reusability & .14
 & 1 & 0 & 1 & 1 & 1
 & 2.8 & \myred{0} & 2.8 & 2.8 & 2.8 & 1.12 & 3.36 & \mygreen{.56}\tabularnewline
&
3.3 & Analyzability & .14
 & 1 & 	\textonehalf & 1 & 1 & 0
 & 2.8 & 1.4 & 2.8 & 2.8 & \myred{0} & .84 & 3.08 & .84\tabularnewline
&
3.4 & Modifiability & .14
 & 0 & 0 & 0 & 1 & 0
 & 0 & 0 & 0 & \mygreen{2.8} & 0 & -.56 & 1.68 & 2.24\tabularnewline
&
3.5 & Testability & .19
 & \textonethird & \texttwothirds & 0 & 1 & 1
 & 1.3 & 2.5 & \myred{0} & \mygreen{3.8} & \mygreen{3.8} & .8 & 3.75 & 1.52\tabularnewline
&
3.6 & Interoperability & .19
 & 0 & 0 & 0 & 1 & 0
 & 0 & 0 & 0 & \mygreen{3.8} & 0 & -.76 & 2.28 & \myred{3.04}\tabularnewline
\hline
\rowcolor{LightGray}
\multicolumn{4}{>{\bf\tiny}r}{1.00}&
\multicolumn{5}{>{\bf\tiny}r}{Maintainability} & \multicolumn{1}{>{\bf\tiny}c}{6.9} & \multicolumn{1}{>{\bf\tiny}c}{3.9} & \multicolumn{1}{>{\bf\tiny}c}{5.6} & \multicolumn{1}{>{\bf\scriptsize}c}{\textcolor{mygreen}{20}} & \multicolumn{1}{>{\bf\tiny}c}{8.6} &&& \multicolumn{1}{>{\bf\tiny}c}{11} \tabularnewline 
\hline
\multirow{4}{*}{\begin{sideways}Portability\end{sideways}}&
4.1 & Adaptability & .28
& 1 & 1 & 	\textonehalf & \textthreequarters & \textthreequarters 
& \mygreen{5.6} & \mygreen{5.6} & \myred{2.8} & 4.2 & 4.2 & 3.43 & 5.52 & 1.12\tabularnewline
&
4.2 & Installability & .28
& 0 & 1 & 0 & 1 & 0
& 0 & \mygreen{5.6} & 0 & \mygreen{5.6} & 0 & -.5 & 4.98 & \myred{3.36}\tabularnewline
&
4.3.1 & Replaceability~(lang) & .16
& 0 & 0 & 0 & 1 & 0
& 0 & 0 & 0 & \mygreen{3.2} & 0 & -.64 & 1.92 & 2.56\tabularnewline
&
4.3.2 & Replaceability~(inst) & .28
& 0 & 1 & \texttwothirds & 1 & 1
& \myred{0} & 5.6 & 3.8 & 5.6 & 5.6 & 1.93 & 6.28 & 1.48\tabularnewline
\hline
\rowcolor{LightGray}
\multicolumn{4}{>{\bf\tiny}r}{1.00}&
\multicolumn{5}{>{\bf\tiny}r}{Portability} & \multicolumn{1}{>{\bf\scriptsize}c}{\textcolor{myred}{5.6}} & \multicolumn{1}{>{\bf\scriptsize}c}{\textcolor{mygreen}{16.8}} & \multicolumn{1}{>{\bf\tiny}c}{ 6.6 } & \multicolumn{1}{>{\bf\scriptsize}c}{\textcolor{mygreen}{18.6}} & \multicolumn{1}{>{\bf\tiny}c}{9.8} &&& \multicolumn{1}{>{\bf\tiny}c}{8.52} \tabularnewline 
\hline
\multirow{5}{*}{\begin{sideways}Security\end{sideways}}&
5.1 & Confidentiality & .15
& 1 & \textonehalf & 0 & 0 & 1
& \mygreen{3} & 1.5 & \myred{0} & \myred{0} & \mygreen{3} & .15 & 2.84 & 1.5\tabularnewline
&
5.2 & Integrity & .25
& \textonehalf & 1 & 0 & \textonehalf & 0
& 2.5 & \mygreen{5} & \myred{0} & 2.5 & \myred{0} & .12 & 3.87 & \myred{3}\tabularnewline
&
5.3 & Non-repudiation & .25
& \textonequarter & 0 & 0 & \textonequarter & 0
& \mygreen{1.2} & 0 & 0 & \mygreen{1.2} & 0 & -.11 & 1.11 & \myred{4.5}\tabularnewline
&
5.4 & Accountability & .15
& 0 & 1 & 0 & 0 & 1
& 0 & \mygreen{3} & 0 & 0 & \mygreen{3} & -.26 & 2.66 & 1.8\tabularnewline
&
5.5 & Authenticity & .20
& \textthreequarters & \textonehalf & 1 & 0 & 1
& 3 & 2 & 4 & \myred{0} & 4 & 1.1 & 4.09 & 1.4\tabularnewline
\hline
\rowcolor{LightGray}
\multicolumn{4}{>{\bf\tiny}r}{1.00}&
\multicolumn{5}{>{\bf\tiny}r}{Security} & \multicolumn{1}{>{\bf\tiny}c}{9.7} & \multicolumn{1}{>{\bf\scriptsize}c}{\textcolor{mygreen}{11.5}} & \multicolumn{1}{>{\bf\scriptsize}c}{\textcolor{myred}{4}} & \multicolumn{1}{>{\bf\scriptsize}c}{\textcolor{myred}{3.7}} & \multicolumn{1}{>{\bf\tiny}c}{10} &&& \multicolumn{1}{>{\bf\scriptsize}c}{\textcolor{myred}{12.2}} \tabularnewline 
\end{tabular}}
\end{center}
\end{table*}

%% file: tab/tab_product.tex
\begin{table}
\begin{center}
\definecolor{LightGray}{gray}{.9}
\caption{Anonymized profile of the five products used for evaluation}\label{table:products}
\centering{\footnotesize
\begin{tabular}{|l|>{\columncolor{colorA}\color{white}\scriptsize}c>{\columncolor{colorB}\scriptsize}c>{\columncolor{colorC}\scriptsize}c>{\columncolor{colorD}\scriptsize}c>{\columncolor{colorE}\scriptsize}c|}\hline
\rowcolor{LightGray}
Metric~$\backslash$~Product	& \ProdA & \ProdB  & \ProdC  & \ProdD & \ProdE \\\hline
No. of users		& $\sim$10 & $\sim$1,000  & $\sim$2,000  & $\sim$300 & $\sim$60 \\
Trans./min	& $\sim$100,000 & $\sim$10  & $\sim$10  & $\sim$200 & $\sim$20 \\
No. of Interfaces	& $\sim$10 & $\sim$30  & $\sim$10  & $\sim$20 & $\sim$4 \\
\hline
\end{tabular}}
\end{center}
\end{table}

%% file: tex/results.tex
\input{tab/tab_modifications}

A 5-hour interview was held with each of the architects of the individual products,
    in order to understand the product's architectures in depth
    and fill out, for each product, the quality compliance questionnaire.
The structure of the interviews
	was based on common architecture review practices.
The collected data is displayed in the middle section of \Tref{QResults},
and the weighted scores are listed in the right-hand section of that table. 
Exceptional values that are more than
	one standard deviation away from the mean are marked
	in \colorbox{myred}{\color{white}\bf red} and
	in \colorbox{yourgreen}{\color{black}\bf green} 
	to emphasize low and high quality results,
	respectively.



\subsection{Results}
The data revealed that the product line suffers major risks
in features
{\scriptsize\ISO{2.1.2}},
{\scriptsize\ISO{2.2.2}},
{\scriptsize\ISO{3.1}},
{\scriptsize\ISO{3.6}},
{\scriptsize\ISO{4.2}},
{\scriptsize\ISO{5.2}}, and
{\scriptsize\ISO{5.3}},
as they have very low values across the product line~(\(\Delta_{\aFeature}>2.7\)).
Fixing these features should achieve the highest quality improvement for the product line.

\Tref{Modifications} lists the $10$ possible architecture modifications 
    that were independently proposed by the expert architects
    for addressing these specific low-valued features.
These modifications are known to achieve
	perfect answers to the questions in \Tref{Questions}, 
	i.e., modifications that could potentially bring about
	full adherence to these quality features.

The total cost, in workdays,
of each modification
is the cost estimation
for applying the change
to the core architecture
or for producing a shared asset,
plus the cost estimation for applying the change
to each of the affected products separately.
The total gain of each modification
is the sum of quality improvements
in each quality feature
in each of the affected products.

For example,
the cost of~$\MI_{8}$
is estimated as 15~workdays
to create the shared access control asset
plus 2~days to implement it in each of the 5~products,
for a total of 25~workdays.
The gain is
2.5 for~\ProdA,
0 for~\ProdB,
5 for~\ProdC,
2.5 for~\ProdD,
and 5 for~\ProdE, 
for a total of~15.


A set of $10$ possible modification spans a search space of size $2^{10}-1$ for the optimal architecture.
Note, however, that some of the modifications in \Tref{Modifications} are mutually incompatible.
For example, different approaches to achieving the very same quality adherence
cannot be implemented together. 
Three out of eight combination of $\{\Mi_1, \Mi_2, \Mi_3\}$ are incompatible, 
and one out of four combinations is incompatible for each of the pairs $\{\Mi_7,\Mi_8\}$ and $\{\Mi_9,\Mi_{10}\}$.
The size of the search space is therefore reduced to: 
\( 2^{10}\times\left(\nicefrac{5}{8}\right)\times\left(\nicefrac{3}{4}\right)^{2}-1=359 \)

Based on the quality gain and cost per modification found in \Tref{Modifications},
\Fref{CaseStudySelection} depicts the quality adherence and total 
cost for each of the $359$ possible modifications subsets, sorted by gain. 
The dotted horizontal bar in \Fref{CaseStudySelection} shows that if the product line management set a cost limit of $\TCR=250$,
the $351^{st}$ set, $\MI_{3,5,6,8,10}$, yields the highest gain of $102.5$
for a cost of $233 <\TCR$.
If the product line management sets \(\gamma=1.6\),
the $201^{st}$ set, $\MI_{2,6,8,10}$, is optimal,
yielding a lower gain of $69.3$ but at a much lower cost of $109$.
Eventually, the management decided they can afford a higher investment,
and 
$\MI_{3,5,6,8,10}$ was selected as the optimal set of modifications.
Applying those modifications to the product line architecture gives 
the highest possible improvement (from~$262.4$ to~$364.9$) 
in adherence to the requirements.

After generating modifications for bridging product line gaps, 
	more modifications were generated for the specific product gaps. 
For each product~$\aProduct$, features with 
	$\weightedVal < \mean{\aFeature} - \stddev{\aFeature}$ 
were analyzed for modifications that would bring their score up to~$1$.
\input{fig_casestudy}

\input{fig_views}

%% file: tab/tab_modifications.tex
\begin{table*}\centering
\begin{center}
\caption{Architecture Modification Options}
\label{table:Modifications}
{\centering
\begin{tabular}{|c|@{}l@{}|@{}l@{}|@{}c@{}c@{}|
>{\columncolor{colorA}\color{white}}c
>{\columncolor{colorB}}c
>{\columncolor{colorC}}c
>{\columncolor{colorD}}c
>{\columncolor{colorE}}c
|cc|}
\hline
\rowcolor{LightGray}
No. & Feature & Modification & Asset & Mutex. & \ProdA & \ProdB & \ProdC & \ProdD & \ProdE & Gain & Cost \tabularnewline
\hline
\(\mu_1\)
 & \ISO{\scriptsize 2.1.2}
 & Monitor latency and throughput metrics
 & Core
 & \(\mu_3\)
 & 3
 & 3
 & 3
 & 3
 & 3
 & 15
 & 55
\tabularnewline
\(\mu_2\)
 & \ISO{\scriptsize 2.2.2}
 & Monitor resource utilization metrics
 & Core
 & \(\mu_3\)
 & 3
 & 3
 & 3
 & 3
 & 3
 & 15
 & 20
\tabularnewline
\(\mu_3\)
 & {\scriptsize\ISO{\scriptsize 2.1.2}/\ISO{\scriptsize 2.2.2}/\ISO{\scriptsize 1.1}}
 & Monitor performance and reliability metrics
 & Core
 & \(\mu_1\),\(\mu_2\)
 & 6
 & 6.7
 & 7.4
 & 6.7
 & 6
 & 33
 & 80
\tabularnewline
\(\mu_4\)
 & \ISO{\scriptsize 3.1}
 & Refine microservices composition
 & Particular
 & -
 & 4
 & 4
 & 4
 & 0
 & 2
 & 14
 & 60
\tabularnewline
\(\mu_5\)
 & \ISO{\scriptsize 3.6}
 & Modernize communication protocols technology
 & Particular
 & -
 & 3.8
 & 3.8
 & 3.8
 & 0
 & 3.8
 & 15.2
 & 64
\tabularnewline
\(\mu_6\)
 & \ISO{\scriptsize 4.2}
 & Containerize services
 & Particular
 & -
 & 5.6
 & 0
 & 5.6
 & 0
 & 5.6
 & 16.8
 & 36
\tabularnewline
\(\mu_7\)
 & \ISO{\scriptsize 5.2}
 & Implement internal access control mechanisms
 & Particular
 & \(\mu_8\)
 & 2.5
 & 0
 & 5
 & 2.5
 & 5
 & 15
 & 40
\tabularnewline
\(\mu_8\)
 & \ISO{\scriptsize 5.2}
 & Implement a common access control asset
 & Shared
 & \(\mu_7\)
 & 2.5
 & 0
 & 5
 & 2.5
 & 5
 & 15
 & 25
\tabularnewline
\(\mu_9\)
 & \ISO{\scriptsize 5.3}
 & Implement internal auditing mechanisms
 & Particular
 & \(\mu_{10}\)
 & 3.7
 & 5
 & 5
 & 3.7
 & 5
 & 22.5
 & 30
\tabularnewline
\(\mu_{10}\)
 & \ISO{\scriptsize 5.3}
 & Implement a common auditing asset
 & Shared
 & \(\mu_9\)
 & 3.7
 & 5
 & 5
 & 3.7
 & 5
 & 22.5
 & 28
\tabularnewline
\hline
\end{tabular}
}
\end{center}
\end{table*}

%% file: fig/fig_casestudy.tex
\begin{figure}[t]
    \centering
    \input{case_study_results-multiplot} 
    \caption{$359$ Modification Subsets}
    \label{fig:CaseStudySelection}
\end{figure}
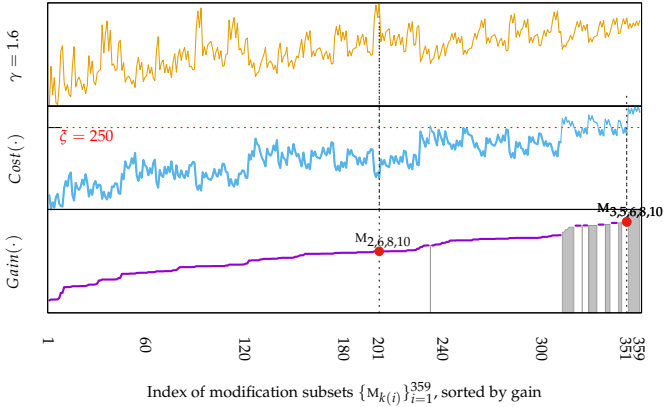

%% file: plot/case_study_results-multiplot.tex
\begin{tikzpicture}[gnuplot]
\path (0.000,0.000) rectangle (9.000,5.500);
\gpcolor{color=gp lt color border}
\gpsetlinetype{gp lt border}
\gpsetdashtype{gp dt solid}
\gpsetlinewidth{1.00}
\draw[gp path] (0.552,5.499)--(0.552,4.125)--(8.447,4.125)--(8.447,5.499)--cycle;
\gpcolor{rgb color={0.000,0.000,0.000}}
\gpsetlinetype{gp lt axes}
\gpsetdashtype{gp dt axes}
\draw[gp path](4.960,4.125)--(4.960,5.499);
\gpcolor{color=gp lt color border}
\node[gp node center,rotate=-270,font={\fontsize{6.0pt}{7.2pt}\selectfont}] at (0.076,4.812) {$\gamma=1.6$};
\gpcolor{rgb color={0.902,0.624,0.000}}
\gpsetlinetype{gp lt border}
\gpsetdashtype{gp dt 1}
\draw[gp path] (0.574,4.125)--(0.596,4.653)--(0.618,4.277)--(0.640,4.503)--(0.662,4.174)%
  --(0.684,4.141)--(0.706,4.402)--(0.727,4.861)--(0.749,4.929)--(0.771,4.441)--(0.793,4.333)%
  --(0.815,4.409)--(0.837,4.276)--(0.859,4.253)--(0.881,4.661)--(0.903,4.915)--(0.925,4.381)%
  --(0.947,4.471)--(0.969,4.509)--(0.991,4.449)--(1.013,4.344)--(1.034,4.419)--(1.056,4.288)%
  --(1.078,4.396)--(1.100,4.795)--(1.122,4.560)--(1.144,4.722)--(1.166,4.451)--(1.188,4.406)%
  --(1.210,4.565)--(1.232,4.594)--(1.254,4.610)--(1.276,5.205)--(1.298,4.832)--(1.320,5.086)%
  --(1.342,5.259)--(1.363,4.860)--(1.385,5.131)--(1.407,4.668)--(1.429,4.686)--(1.451,4.600)%
  --(1.473,4.615)--(1.495,4.966)--(1.517,4.999)--(1.539,4.602)--(1.561,4.703)--(1.583,4.444)%
  --(1.605,4.502)--(1.627,4.524)--(1.649,4.490)--(1.670,4.418)--(1.692,4.470)--(1.714,4.374)%
  --(1.736,4.660)--(1.758,4.774)--(1.780,4.610)--(1.802,4.707)--(1.824,4.453)--(1.846,4.511)%
  --(1.868,4.533)--(1.890,4.675)--(1.912,4.561)--(1.934,4.643)--(1.956,4.495)--(1.977,4.466)%
  --(1.999,4.869)--(2.021,5.048)--(2.043,4.610)--(2.065,4.702)--(2.087,4.738)--(2.109,4.569)%
  --(2.131,4.680)--(2.153,4.569)--(2.175,4.649)--(2.197,4.504)--(2.219,4.707)--(2.241,4.822)%
  --(2.263,4.577)--(2.285,4.786)--(2.306,4.885)--(2.328,4.734)--(2.350,4.843)--(2.372,4.904)%
  --(2.394,4.747)--(2.416,4.859)--(2.438,4.648)--(2.460,4.658)--(2.482,4.608)--(2.504,4.618)%
  --(2.526,5.142)--(2.548,5.390)--(2.570,5.170)--(2.592,5.431)--(2.613,4.794)--(2.635,4.916)%
  --(2.657,4.809)--(2.679,4.935)--(2.701,4.965)--(2.723,4.985)--(2.745,4.887)--(2.767,4.739)%
  --(2.789,4.845)--(2.811,4.904)--(2.833,4.752)--(2.855,4.861)--(2.877,4.655)--(2.899,4.665)%
  --(2.921,4.809)--(2.942,4.824)--(2.964,5.272)--(2.986,5.013)--(3.008,5.196)--(3.030,5.304)%
  --(3.052,5.034)--(3.074,5.226)--(3.096,4.875)--(3.118,4.891)--(3.140,4.813)--(3.162,4.827)%
  --(3.184,5.011)--(3.206,5.031)--(3.228,4.670)--(3.249,4.742)--(3.271,4.636)--(3.293,4.702)%
  --(3.315,4.519)--(3.337,4.564)--(3.359,4.581)--(3.381,4.677)--(3.403,4.749)--(3.425,4.806)%
  --(3.447,4.903)--(3.469,4.640)--(3.491,4.703)--(3.513,4.727)--(3.535,4.690)--(3.556,4.611)%
  --(3.578,4.668)--(3.600,4.561)--(3.622,4.861)--(3.644,4.967)--(3.666,4.710)--(3.688,4.784)%
  --(3.710,4.812)--(3.732,4.906)--(3.754,4.648)--(3.776,4.710)--(3.798,4.734)--(3.820,4.619)%
  --(3.842,4.718)--(3.864,4.790)--(3.885,4.768)--(3.907,4.904)--(3.929,5.010)--(3.951,4.774)%
  --(3.973,4.988)--(3.995,5.109)--(4.017,5.003)--(4.039,5.127)--(4.061,4.782)--(4.083,4.860)%
  --(4.105,4.792)--(4.127,4.871)--(4.149,4.889)--(4.171,4.901)--(4.192,4.842)--(4.214,4.745)%
  --(4.236,4.816)--(4.258,4.853)--(4.280,4.754)--(4.302,4.826)--(4.324,4.685)--(4.346,4.692)%
  --(4.368,5.053)--(4.390,5.185)--(4.412,5.069)--(4.434,5.206)--(4.456,4.990)--(4.478,5.108)%
  --(4.500,5.005)--(4.521,5.126)--(4.543,4.788)--(4.565,4.865)--(4.587,4.798)--(4.609,4.876)%
  --(4.631,4.894)--(4.653,4.905)--(4.675,5.066)--(4.697,4.926)--(4.719,5.028)--(4.741,5.083)%
  --(4.763,4.938)--(4.785,5.043)--(4.807,4.841)--(4.828,4.851)--(4.850,4.801)--(4.872,4.810)%
  --(4.894,5.283)--(4.916,5.470)--(4.938,5.306)--(4.960,5.499)--(4.982,4.983)--(5.004,5.094)%
  --(5.026,4.996)--(5.048,5.110)--(5.070,5.136)--(5.092,5.154)--(5.114,4.930)--(5.135,4.942)%
  --(5.157,5.067)--(5.179,4.930)--(5.201,5.030)--(5.223,5.083)--(5.245,4.942)--(5.267,5.044)%
  --(5.289,4.847)--(5.311,4.857)--(5.333,5.094)--(5.355,5.227)--(5.377,5.111)--(5.399,5.247)%
  --(5.421,4.934)--(5.443,4.946)--(5.464,5.178)--(5.486,5.195)--(5.508,4.714)--(5.530,4.768)%
  --(5.552,4.854)--(5.574,4.926)--(5.596,4.818)--(5.618,4.886)--(5.640,4.692)--(5.662,4.742)%
  --(5.684,4.760)--(5.706,4.860)--(5.728,4.932)--(5.750,4.749)--(5.771,4.805)--(5.793,4.891)%
  --(5.815,4.965)--(5.837,4.796)--(5.859,4.897)--(5.881,4.970)--(5.903,5.002)--(5.925,5.089)%
  --(5.947,5.013)--(5.969,5.102)--(5.991,4.960)--(6.013,5.040)--(6.035,4.970)--(6.057,5.051)%
  --(6.079,4.811)--(6.100,4.869)--(6.122,4.818)--(6.144,4.878)--(6.166,4.891)--(6.188,4.899)%
  --(6.210,5.007)--(6.232,5.092)--(6.254,5.017)--(6.276,5.105)--(6.298,5.158)--(6.320,5.268)%
  --(6.342,5.172)--(6.364,5.284)--(6.386,4.959)--(6.407,5.036)--(6.429,4.968)--(6.451,5.047)%
  --(6.473,5.064)--(6.495,5.076)--(6.517,5.018)--(6.539,4.921)--(6.561,4.992)--(6.583,5.029)%
  --(6.605,4.930)--(6.627,5.003)--(6.649,4.859)--(6.671,4.867)--(6.693,5.218)--(6.714,5.337)%
  --(6.736,5.232)--(6.758,5.354)--(6.780,5.040)--(6.802,5.128)--(6.824,5.051)--(6.846,5.140)%
  --(6.868,5.160)--(6.890,5.268)--(6.912,5.173)--(6.934,5.284)--(6.956,4.964)--(6.978,5.040)%
  --(7.000,4.973)--(7.022,5.051)--(7.043,5.068)--(7.065,5.079)--(7.087,4.927)--(7.109,4.936)%
  --(7.131,5.044)--(7.153,5.130)--(7.175,5.055)--(7.197,5.143)--(7.219,5.102)--(7.241,5.114)%
  --(7.263,5.256)--(7.285,5.375)--(7.307,5.271)--(7.329,5.392)--(7.350,5.105)--(7.372,5.117)%
  --(7.394,4.880)--(7.416,4.937)--(7.438,4.914)--(7.460,4.971)--(7.482,5.004)--(7.504,5.069)%
  --(7.526,5.013)--(7.548,5.079)--(7.570,5.165)--(7.592,5.249)--(7.614,5.176)--(7.636,5.261)%
  --(7.658,5.124)--(7.679,5.202)--(7.701,5.134)--(7.723,5.213)--(7.745,4.973)--(7.767,5.033)%
  --(7.789,4.981)--(7.811,5.042)--(7.833,5.054)--(7.855,5.063)--(7.877,5.169)--(7.899,5.252)%
  --(7.921,5.179)--(7.943,5.264)--(7.965,5.038)--(7.986,5.104)--(8.008,5.047)--(8.030,5.114)%
  --(8.052,5.201)--(8.074,5.285)--(8.096,5.211)--(8.118,5.297)--(8.140,5.089)--(8.162,5.098)%
  --(8.184,5.204)--(8.206,5.287)--(8.228,5.215)--(8.250,5.299)--(8.272,5.158)--(8.293,5.223)%
  --(8.315,5.167)--(8.337,5.233)--(8.359,5.191)--(8.381,5.256)--(8.403,5.199)--(8.425,5.266);
\gpcolor{color=gp lt color border}
\gpsetdashtype{gp dt solid}
\draw[gp path] (0.552,5.499)--(0.552,4.125)--(8.447,4.125)--(8.447,5.499)--cycle;
\gpdefrectangularnode{gp plot 1}{\pgfpoint{0.552cm}{4.125cm}}{\pgfpoint{8.447cm}{5.499cm}}
\draw[gp path] (0.552,3.840)--(0.732,3.840);
\gpcolor{rgb color={1.000,0.000,0.000}}
\node[gp node right,font={\fontsize{6.0pt}{7.2pt}\selectfont}] at (1.552,3.758) {${\TCR}=250$};
\gpcolor{color=gp lt color border}
\draw[gp path] (0.552,4.124)--(0.552,2.750)--(8.447,2.750)--(8.447,4.124)--cycle;
\gpcolor{rgb color={1.000,0.000,0.000}}
\gpsetlinetype{gp lt axes}
\gpsetdashtype{gp dt axes}
\draw[gp path](0.552,3.840)--(8.447,3.840);
\gpcolor{rgb color={0.000,0.000,0.000}}
\draw[gp path](4.960,2.750)--(4.960,4.188);
\draw[gp path](8.250,2.750)--(8.250,3.840);
\gpcolor{color=gp lt color border}
\node[gp node center,rotate=-270,font={\fontsize{6.0pt}{7.2pt}\selectfont}] at (0.103,3.437) {$Cost(\cdot)$};
\gpcolor{rgb color={0.337,0.706,0.914}}
\gpsetlinetype{gp lt border}
\gpsetdashtype{gp dt solid}
\draw[gp path] (0.574,2.940)--(0.596,2.750)--(0.618,2.845)--(0.640,2.774)--(0.662,2.916)%
  --(0.684,2.958)--(0.706,2.826)--(0.727,2.797)--(0.749,2.788)--(0.771,3.034)--(0.793,3.129)%
  --(0.815,3.058)--(0.837,3.200)--(0.859,3.243)--(0.881,2.940)--(0.903,2.868)--(0.925,3.105)%
  --(0.947,3.034)--(0.969,3.011)--(0.991,3.053)--(1.013,3.148)--(1.034,3.077)--(1.056,3.219)%
  --(1.078,3.110)--(1.100,2.921)--(1.122,3.015)--(1.144,2.944)--(1.166,3.086)--(1.188,3.129)%
  --(1.210,3.034)--(1.232,3.082)--(1.254,3.072)--(1.276,2.892)--(1.298,2.987)--(1.320,2.916)%
  --(1.342,2.883)--(1.363,2.977)--(1.385,2.906)--(1.407,3.058)--(1.429,3.048)--(1.451,3.101)%
  --(1.473,3.091)--(1.495,2.968)--(1.517,2.958)--(1.539,3.224)--(1.561,3.153)--(1.583,3.390)%
  --(1.605,3.319)--(1.627,3.295)--(1.649,3.338)--(1.670,3.432)--(1.692,3.361)--(1.714,3.503)%
  --(1.736,3.200)--(1.758,3.129)--(1.780,3.243)--(1.802,3.172)--(1.824,3.409)--(1.846,3.338)%
  --(1.868,3.314)--(1.890,3.205)--(1.912,3.300)--(1.934,3.229)--(1.956,3.371)--(1.977,3.413)%
  --(1.999,3.110)--(2.021,3.039)--(2.043,3.276)--(2.065,3.205)--(2.087,3.181)--(2.109,3.319)%
  --(2.131,3.224)--(2.153,3.319)--(2.175,3.247)--(2.197,3.390)--(2.219,3.224)--(2.241,3.153)%
  --(2.263,3.338)--(2.285,3.205)--(2.306,3.176)--(2.328,3.271)--(2.350,3.200)--(2.372,3.167)%
  --(2.394,3.262)--(2.416,3.191)--(2.438,3.342)--(2.460,3.333)--(2.482,3.385)--(2.504,3.375)%
  --(2.526,3.082)--(2.548,3.011)--(2.570,3.072)--(2.592,3.001)--(2.613,3.247)--(2.635,3.176)%
  --(2.657,3.238)--(2.679,3.167)--(2.701,3.153)--(2.723,3.143)--(2.745,3.195)--(2.767,3.290)%
  --(2.789,3.219)--(2.811,3.186)--(2.833,3.281)--(2.855,3.210)--(2.877,3.361)--(2.899,3.352)%
  --(2.921,3.252)--(2.942,3.243)--(2.964,3.063)--(2.986,3.157)--(3.008,3.086)--(3.030,3.053)%
  --(3.052,3.148)--(3.074,3.077)--(3.096,3.229)--(3.118,3.219)--(3.140,3.271)--(3.162,3.262)%
  --(3.184,3.176)--(3.206,3.167)--(3.228,3.484)--(3.249,3.413)--(3.271,3.527)--(3.293,3.456)%
  --(3.315,3.693)--(3.337,3.622)--(3.359,3.598)--(3.381,3.503)--(3.403,3.432)--(3.425,3.394)%
  --(3.447,3.323)--(3.469,3.560)--(3.491,3.489)--(3.513,3.465)--(3.535,3.508)--(3.556,3.603)%
  --(3.578,3.532)--(3.600,3.674)--(3.622,3.371)--(3.644,3.300)--(3.666,3.508)--(3.688,3.437)%
  --(3.710,3.413)--(3.732,3.342)--(3.754,3.579)--(3.776,3.508)--(3.798,3.484)--(3.820,3.622)%
  --(3.842,3.527)--(3.864,3.456)--(3.885,3.489)--(3.907,3.394)--(3.929,3.323)--(3.951,3.508)%
  --(3.973,3.366)--(3.995,3.295)--(4.017,3.356)--(4.039,3.285)--(4.061,3.532)--(4.083,3.461)%
  --(4.105,3.522)--(4.127,3.451)--(4.149,3.437)--(4.171,3.428)--(4.192,3.480)--(4.214,3.574)%
  --(4.236,3.503)--(4.258,3.470)--(4.280,3.565)--(4.302,3.494)--(4.324,3.645)--(4.346,3.636)%
  --(4.368,3.342)--(4.390,3.271)--(4.412,3.333)--(4.434,3.262)--(4.456,3.385)--(4.478,3.314)%
  --(4.500,3.375)--(4.521,3.304)--(4.543,3.551)--(4.565,3.480)--(4.587,3.541)--(4.609,3.470)%
  --(4.631,3.456)--(4.653,3.446)--(4.675,3.347)--(4.697,3.442)--(4.719,3.371)--(4.741,3.338)%
  --(4.763,3.432)--(4.785,3.361)--(4.807,3.513)--(4.828,3.503)--(4.850,3.555)--(4.872,3.546)%
  --(4.894,3.252)--(4.916,3.181)--(4.938,3.243)--(4.960,3.172)--(4.982,3.418)--(5.004,3.347)%
  --(5.026,3.409)--(5.048,3.338)--(5.070,3.323)--(5.092,3.314)--(5.114,3.461)--(5.135,3.451)%
  --(5.157,3.366)--(5.179,3.461)--(5.201,3.390)--(5.223,3.356)--(5.245,3.451)--(5.267,3.380)%
  --(5.289,3.532)--(5.311,3.522)--(5.333,3.366)--(5.355,3.295)--(5.377,3.356)--(5.399,3.285)%
  --(5.421,3.480)--(5.443,3.470)--(5.464,3.347)--(5.486,3.338)--(5.508,3.788)--(5.530,3.717)%
  --(5.552,3.655)--(5.574,3.584)--(5.596,3.698)--(5.618,3.627)--(5.640,3.863)--(5.662,3.792)%
  --(5.684,3.769)--(5.706,3.674)--(5.728,3.603)--(5.750,3.811)--(5.771,3.740)--(5.793,3.679)%
  --(5.815,3.608)--(5.837,3.792)--(5.859,3.698)--(5.881,3.627)--(5.903,3.627)--(5.925,3.555)%
  --(5.947,3.617)--(5.969,3.546)--(5.991,3.669)--(6.013,3.598)--(6.035,3.660)--(6.057,3.589)%
  --(6.079,3.835)--(6.100,3.764)--(6.122,3.826)--(6.144,3.754)--(6.166,3.740)--(6.188,3.731)%
  --(6.210,3.645)--(6.232,3.574)--(6.254,3.636)--(6.276,3.565)--(6.298,3.536)--(6.320,3.465)%
  --(6.342,3.527)--(6.364,3.456)--(6.386,3.702)--(6.407,3.631)--(6.429,3.693)--(6.451,3.622)%
  --(6.473,3.608)--(6.495,3.598)--(6.517,3.650)--(6.539,3.745)--(6.561,3.674)--(6.583,3.641)%
  --(6.605,3.735)--(6.627,3.664)--(6.649,3.816)--(6.671,3.807)--(6.693,3.513)--(6.714,3.442)%
  --(6.736,3.503)--(6.758,3.432)--(6.780,3.650)--(6.802,3.579)--(6.824,3.641)--(6.846,3.570)%
  --(6.868,3.555)--(6.890,3.484)--(6.912,3.546)--(6.934,3.475)--(6.956,3.721)--(6.978,3.650)%
  --(7.000,3.712)--(7.022,3.641)--(7.043,3.627)--(7.065,3.617)--(7.087,3.764)--(7.109,3.754)%
  --(7.131,3.669)--(7.153,3.598)--(7.175,3.660)--(7.197,3.589)--(7.219,3.631)--(7.241,3.622)%
  --(7.263,3.536)--(7.285,3.465)--(7.307,3.527)--(7.329,3.456)--(7.350,3.650)--(7.372,3.641)%
  --(7.394,3.958)--(7.416,3.887)--(7.438,3.982)--(7.460,3.911)--(7.482,3.930)--(7.504,3.859)%
  --(7.526,3.920)--(7.548,3.849)--(7.570,3.797)--(7.592,3.726)--(7.614,3.788)--(7.636,3.717)%
  --(7.658,3.840)--(7.679,3.769)--(7.701,3.830)--(7.723,3.759)--(7.745,4.006)--(7.767,3.934)%
  --(7.789,3.996)--(7.811,3.925)--(7.833,3.911)--(7.855,3.901)--(7.877,3.816)--(7.899,3.745)%
  --(7.921,3.807)--(7.943,3.735)--(7.965,3.953)--(7.986,3.882)--(8.008,3.944)--(8.030,3.873)%
  --(8.052,3.821)--(8.074,3.750)--(8.096,3.811)--(8.118,3.740)--(8.140,3.934)--(8.162,3.925)%
  --(8.184,3.840)--(8.206,3.769)--(8.228,3.830)--(8.250,3.759)--(8.272,4.100)--(8.293,4.029)%
  --(8.315,4.091)--(8.337,4.020)--(8.359,4.124)--(8.381,4.053)--(8.403,4.115)--(8.425,4.043);
\draw[gp path] (7.394,3.958)--(7.416,3.887)--(7.438,3.982)--(7.460,3.911)--(7.482,3.930)%
  --(7.504,3.859)--(7.526,3.920)--(7.548,3.849);
\draw[gp path] (7.745,4.006)--(7.767,3.934)--(7.789,3.996)--(7.811,3.925)--(7.833,3.911)%
  --(7.855,3.901);
\draw[gp path] (7.965,3.953)--(7.986,3.882)--(8.008,3.944)--(8.030,3.873);
\draw[gp path] (8.140,3.934)--(8.162,3.925)--(8.184,3.840);
\draw[gp path] (8.272,4.100)--(8.293,4.029)--(8.315,4.091)--(8.337,4.020)--(8.359,4.124)%
  --(8.381,4.053)--(8.403,4.115)--(8.425,4.043);
\gpsetdashtype{gp dt 1}
\gpsetlinewidth{2.00}
\draw[gp path] (0.574,2.940)--(0.596,2.750)--(0.618,2.845)--(0.640,2.774)--(0.662,2.916)%
  --(0.684,2.958)--(0.706,2.826)--(0.727,2.797)--(0.749,2.788)--(0.771,3.034)--(0.793,3.129)%
  --(0.815,3.058)--(0.837,3.200)--(0.859,3.243)--(0.881,2.940)--(0.903,2.868)--(0.925,3.105)%
  --(0.947,3.034)--(0.969,3.011)--(0.991,3.053)--(1.013,3.148)--(1.034,3.077)--(1.056,3.219)%
  --(1.078,3.110)--(1.100,2.921)--(1.122,3.015)--(1.144,2.944)--(1.166,3.086)--(1.188,3.129)%
  --(1.210,3.034)--(1.232,3.082)--(1.254,3.072)--(1.276,2.892)--(1.298,2.987)--(1.320,2.916)%
  --(1.342,2.883)--(1.363,2.977)--(1.385,2.906)--(1.407,3.058)--(1.429,3.048)--(1.451,3.101)%
  --(1.473,3.091)--(1.495,2.968)--(1.517,2.958)--(1.539,3.224)--(1.561,3.153)--(1.583,3.390)%
  --(1.605,3.319)--(1.627,3.295)--(1.649,3.338)--(1.670,3.432)--(1.692,3.361)--(1.714,3.503)%
  --(1.736,3.200)--(1.758,3.129)--(1.780,3.243)--(1.802,3.172)--(1.824,3.409)--(1.846,3.338)%
  --(1.868,3.314)--(1.890,3.205)--(1.912,3.300)--(1.934,3.229)--(1.956,3.371)--(1.977,3.413)%
  --(1.999,3.110)--(2.021,3.039)--(2.043,3.276)--(2.065,3.205)--(2.087,3.181)--(2.109,3.319)%
  --(2.131,3.224)--(2.153,3.319)--(2.175,3.247)--(2.197,3.390)--(2.219,3.224)--(2.241,3.153)%
  --(2.263,3.338)--(2.285,3.205)--(2.306,3.176)--(2.328,3.271)--(2.350,3.200)--(2.372,3.167)%
  --(2.394,3.262)--(2.416,3.191)--(2.438,3.342)--(2.460,3.333)--(2.482,3.385)--(2.504,3.375)%
  --(2.526,3.082)--(2.548,3.011)--(2.570,3.072)--(2.592,3.001)--(2.613,3.247)--(2.635,3.176)%
  --(2.657,3.238)--(2.679,3.167)--(2.701,3.153)--(2.723,3.143)--(2.745,3.195)--(2.767,3.290)%
  --(2.789,3.219)--(2.811,3.186)--(2.833,3.281)--(2.855,3.210)--(2.877,3.361)--(2.899,3.352)%
  --(2.921,3.252)--(2.942,3.243)--(2.964,3.063)--(2.986,3.157)--(3.008,3.086)--(3.030,3.053)%
  --(3.052,3.148)--(3.074,3.077)--(3.096,3.229)--(3.118,3.219)--(3.140,3.271)--(3.162,3.262)%
  --(3.184,3.176)--(3.206,3.167)--(3.228,3.484)--(3.249,3.413)--(3.271,3.527)--(3.293,3.456)%
  --(3.315,3.693)--(3.337,3.622)--(3.359,3.598)--(3.381,3.503)--(3.403,3.432)--(3.425,3.394)%
  --(3.447,3.323)--(3.469,3.560)--(3.491,3.489)--(3.513,3.465)--(3.535,3.508)--(3.556,3.603)%
  --(3.578,3.532)--(3.600,3.674)--(3.622,3.371)--(3.644,3.300)--(3.666,3.508)--(3.688,3.437)%
  --(3.710,3.413)--(3.732,3.342)--(3.754,3.579)--(3.776,3.508)--(3.798,3.484)--(3.820,3.622)%
  --(3.842,3.527)--(3.864,3.456)--(3.885,3.489)--(3.907,3.394)--(3.929,3.323)--(3.951,3.508)%
  --(3.973,3.366)--(3.995,3.295)--(4.017,3.356)--(4.039,3.285)--(4.061,3.532)--(4.083,3.461)%
  --(4.105,3.522)--(4.127,3.451)--(4.149,3.437)--(4.171,3.428)--(4.192,3.480)--(4.214,3.574)%
  --(4.236,3.503)--(4.258,3.470)--(4.280,3.565)--(4.302,3.494)--(4.324,3.645)--(4.346,3.636)%
  --(4.368,3.342)--(4.390,3.271)--(4.412,3.333)--(4.434,3.262)--(4.456,3.385)--(4.478,3.314)%
  --(4.500,3.375)--(4.521,3.304)--(4.543,3.551)--(4.565,3.480)--(4.587,3.541)--(4.609,3.470)%
  --(4.631,3.456)--(4.653,3.446)--(4.675,3.347)--(4.697,3.442)--(4.719,3.371)--(4.741,3.338)%
  --(4.763,3.432)--(4.785,3.361)--(4.807,3.513)--(4.828,3.503)--(4.850,3.555)--(4.872,3.546)%
  --(4.894,3.252)--(4.916,3.181)--(4.938,3.243)--(4.960,3.172)--(4.982,3.418)--(5.004,3.347)%
  --(5.026,3.409)--(5.048,3.338)--(5.070,3.323)--(5.092,3.314)--(5.114,3.461)--(5.135,3.451)%
  --(5.157,3.366)--(5.179,3.461)--(5.201,3.390)--(5.223,3.356)--(5.245,3.451)--(5.267,3.380)%
  --(5.289,3.532)--(5.311,3.522)--(5.333,3.366)--(5.355,3.295)--(5.377,3.356)--(5.399,3.285)%
  --(5.421,3.480)--(5.443,3.470)--(5.464,3.347)--(5.486,3.338)--(5.508,3.788)--(5.530,3.717)%
  --(5.552,3.655)--(5.574,3.584)--(5.596,3.698)--(5.618,3.627);
\draw[gp path] (5.662,3.792)--(5.684,3.769)--(5.706,3.674)--(5.728,3.603)--(5.750,3.811)%
  --(5.771,3.740)--(5.793,3.679)--(5.815,3.608)--(5.837,3.792)--(5.859,3.698)--(5.881,3.627)%
  --(5.903,3.627)--(5.925,3.555)--(5.947,3.617)--(5.969,3.546)--(5.991,3.669)--(6.013,3.598)%
  --(6.035,3.660)--(6.057,3.589)--(6.079,3.835)--(6.100,3.764)--(6.122,3.826)--(6.144,3.754)%
  --(6.166,3.740)--(6.188,3.731)--(6.210,3.645)--(6.232,3.574)--(6.254,3.636)--(6.276,3.565)%
  --(6.298,3.536)--(6.320,3.465)--(6.342,3.527)--(6.364,3.456)--(6.386,3.702)--(6.407,3.631)%
  --(6.429,3.693)--(6.451,3.622)--(6.473,3.608)--(6.495,3.598)--(6.517,3.650)--(6.539,3.745)%
  --(6.561,3.674)--(6.583,3.641)--(6.605,3.735)--(6.627,3.664)--(6.649,3.816)--(6.671,3.807)%
  --(6.693,3.513)--(6.714,3.442)--(6.736,3.503)--(6.758,3.432)--(6.780,3.650)--(6.802,3.579)%
  --(6.824,3.641)--(6.846,3.570)--(6.868,3.555)--(6.890,3.484)--(6.912,3.546)--(6.934,3.475)%
  --(6.956,3.721)--(6.978,3.650)--(7.000,3.712)--(7.022,3.641)--(7.043,3.627)--(7.065,3.617)%
  --(7.087,3.764)--(7.109,3.754)--(7.131,3.669)--(7.153,3.598)--(7.175,3.660)--(7.197,3.589)%
  --(7.219,3.631)--(7.241,3.622)--(7.263,3.536)--(7.285,3.465)--(7.307,3.527)--(7.329,3.456)%
  --(7.350,3.650)--(7.372,3.641);
\draw[gp path] (7.570,3.797)--(7.592,3.726)--(7.614,3.788)--(7.636,3.717);
\draw[gp path] (7.679,3.769)--(7.701,3.830)--(7.723,3.759);
\draw[gp path] (7.877,3.816)--(7.899,3.745)--(7.921,3.807)--(7.943,3.735);
\draw[gp path] (8.052,3.821)--(8.074,3.750)--(8.096,3.811)--(8.118,3.740);
\draw[gp path] (8.206,3.769)--(8.228,3.830)--(8.250,3.759);
\gpcolor{color=gp lt color border}
\gpsetdashtype{gp dt solid}
\gpsetlinewidth{1.00}
\draw[gp path] (0.552,4.124)--(0.552,2.750)--(8.447,2.750)--(8.447,4.124)--cycle;
\gpdefrectangularnode{gp plot 2}{\pgfpoint{0.552cm}{2.750cm}}{\pgfpoint{8.447cm}{4.124cm}}
\node[gp node left,rotate=-90,font={\fontsize{6.0pt}{7.2pt}\selectfont}] at (0.574,1.191) {$1$};
\node[gp node left,rotate=-90,font={\fontsize{6.0pt}{7.2pt}\selectfont}] at (1.868,1.191) {$60$};
\node[gp node left,rotate=-90,font={\fontsize{6.0pt}{7.2pt}\selectfont}] at (3.184,1.191) {$120$};
\node[gp node left,rotate=-90,font={\fontsize{6.0pt}{7.2pt}\selectfont}] at (4.500,1.191) {$180$};
\node[gp node left,rotate=-90,font={\fontsize{6.0pt}{7.2pt}\selectfont}] at (4.960,1.191) {$201$};
\node[gp node left,rotate=-90,font={\fontsize{6.0pt}{7.2pt}\selectfont}] at (5.815,1.191) {$240$};
\node[gp node left,rotate=-90,font={\fontsize{6.0pt}{7.2pt}\selectfont}] at (7.131,1.191) {$300$};
\node[gp node left,rotate=-90,font={\fontsize{6.0pt}{7.2pt}\selectfont}] at (8.250,1.191) {$351$};
\node[gp node left,rotate=-90,font={\fontsize{6.0pt}{7.2pt}\selectfont}] at (8.425,1.191) {$359$};
\draw[gp path] (0.552,2.750)--(0.552,1.375)--(8.447,1.375)--(8.447,2.750)--cycle;
\gpcolor{rgb color={0.000,0.000,0.000}}
\gpsetlinetype{gp lt axes}
\gpsetdashtype{gp dt axes}
\draw[gp path](4.960,1.375)--(4.960,5.193);
\draw[gp path](8.250,1.375)--(8.250,4.326);
\gpcolor{color=gp lt color border}
\node[gp node center,rotate=-270,font={\fontsize{6.0pt}{7.2pt}\selectfont}] at (0.076,2.062) {$Gain(\cdot)$};
\node[gp node center,font={\fontsize{6.0pt}{7.2pt}\selectfont}] at (4.499,0.361) {Index of modification subsets~$\{{\MI_{k(i)}}\}_{i=1}^{359}$, sorted by gain};
\gpfill{color=gp lt color border,opacity=0.25} (5.640,2.272)--(5.640,1.375)--(5.640,1.375)--cycle;
  \gpfill{color=gp lt color border,opacity=0.25} (7.394,2.449)--(7.416,2.449)--(7.438,2.484)--(7.460,2.484)%
    --(7.482,2.516)--(7.504,2.516)--(7.526,2.516)--(7.548,2.516)--(7.548,1.375)--(7.394,1.375)--cycle;
\gpfill{color=gp lt color border,opacity=0.25} (7.658,2.538)--(7.658,1.375)--(7.658,1.375)--cycle;
\gpfill{color=gp lt color border,opacity=0.25} (7.745,2.538)--(7.767,2.538)--(7.789,2.538)--(7.811,2.538)%
    --(7.833,2.538)--(7.855,2.538)--(7.855,1.375)--(7.745,1.375)--cycle;
\gpfill{color=gp lt color border,opacity=0.25} (7.965,2.552)--(7.986,2.552)--(8.008,2.552)--(8.030,2.552)%
    --(8.030,1.375)--(7.965,1.375)--cycle;
  \gpfill{color=gp lt color border,opacity=0.25} (8.140,2.573)--(8.162,2.573)--(8.184,2.585)--(8.184,1.375)--(8.140,1.375)--cycle;
  \gpfill{color=gp lt color border,opacity=0.25} (8.272,2.715)--(8.293,2.715)--(8.315,2.715)--(8.337,2.715)%
    --(8.359,2.750)--(8.381,2.750)--(8.403,2.750)--(8.425,2.750)--(8.425,1.375)--(8.272,1.375)--cycle;
\gpcolor{rgb color={0.580,0.000,0.827}}
\gpsetlinetype{gp lt border}
\gpsetdashtype{gp dt solid}
\gpsetlinewidth{2.00}
\draw[gp path] (0.574,1.540)--(0.596,1.552)--(0.618,1.552)--(0.640,1.552)--(0.662,1.552)%
  --(0.684,1.554)--(0.706,1.573)--(0.727,1.641)--(0.749,1.641)--(0.771,1.717)--(0.793,1.717)%
  --(0.815,1.717)--(0.837,1.717)--(0.859,1.720)--(0.881,1.729)--(0.903,1.729)--(0.925,1.729)%
  --(0.947,1.729)--(0.969,1.729)--(0.991,1.731)--(1.013,1.731)--(1.034,1.731)--(1.056,1.731)%
  --(1.078,1.739)--(1.100,1.750)--(1.122,1.750)--(1.144,1.750)--(1.166,1.750)--(1.188,1.753)%
  --(1.210,1.764)--(1.232,1.806)--(1.254,1.806)--(1.276,1.818)--(1.298,1.818)--(1.320,1.818)%
  --(1.342,1.818)--(1.363,1.818)--(1.385,1.818)--(1.407,1.818)--(1.429,1.818)--(1.451,1.820)%
  --(1.473,1.820)--(1.495,1.839)--(1.517,1.839)--(1.539,1.894)--(1.561,1.894)--(1.583,1.894)%
  --(1.605,1.894)--(1.627,1.894)--(1.649,1.897)--(1.670,1.897)--(1.692,1.897)--(1.714,1.897)%
  --(1.736,1.906)--(1.758,1.906)--(1.780,1.908)--(1.802,1.908)--(1.824,1.908)--(1.846,1.908)%
  --(1.868,1.908)--(1.890,1.916)--(1.912,1.916)--(1.934,1.916)--(1.956,1.916)--(1.977,1.918)%
  --(1.999,1.927)--(2.021,1.927)--(2.043,1.927)--(2.065,1.927)--(2.087,1.927)--(2.109,1.930)%
  --(2.131,1.930)--(2.153,1.930)--(2.175,1.930)--(2.197,1.930)--(2.219,1.942)--(2.241,1.942)%
  --(2.263,1.944)--(2.285,1.963)--(2.306,1.983)--(2.328,1.983)--(2.350,1.983)--(2.372,1.983)%
  --(2.394,1.983)--(2.416,1.983)--(2.438,1.983)--(2.460,1.983)--(2.482,1.985)--(2.504,1.985)%
  --(2.526,1.995)--(2.548,1.995)--(2.570,1.995)--(2.592,1.995)--(2.613,1.995)--(2.635,1.995)%
  --(2.657,1.995)--(2.679,1.995)--(2.701,1.995)--(2.723,1.995)--(2.745,1.997)--(2.767,1.997)%
  --(2.789,1.997)--(2.811,1.997)--(2.833,1.997)--(2.855,1.997)--(2.877,1.997)--(2.899,1.997)%
  --(2.921,2.004)--(2.942,2.004)--(2.964,2.016)--(2.986,2.016)--(3.008,2.016)--(3.030,2.016)%
  --(3.052,2.016)--(3.074,2.016)--(3.096,2.016)--(3.118,2.016)--(3.140,2.018)--(3.162,2.018)%
  --(3.184,2.030)--(3.206,2.030)--(3.228,2.071)--(3.249,2.071)--(3.271,2.074)--(3.293,2.074)%
  --(3.315,2.074)--(3.337,2.074)--(3.359,2.074)--(3.381,2.086)--(3.403,2.086)--(3.425,2.093)%
  --(3.447,2.093)--(3.469,2.093)--(3.491,2.093)--(3.513,2.093)--(3.535,2.095)--(3.556,2.095)%
  --(3.578,2.095)--(3.600,2.095)--(3.622,2.104)--(3.644,2.104)--(3.666,2.107)--(3.688,2.107)%
  --(3.710,2.107)--(3.732,2.107)--(3.754,2.107)--(3.776,2.107)--(3.798,2.107)--(3.820,2.109)%
  --(3.842,2.121)--(3.864,2.121)--(3.885,2.128)--(3.907,2.140)--(3.929,2.140)--(3.951,2.142)%
  --(3.973,2.160)--(3.995,2.160)--(4.017,2.160)--(4.039,2.160)--(4.061,2.160)--(4.083,2.160)%
  --(4.105,2.160)--(4.127,2.160)--(4.149,2.160)--(4.171,2.160)--(4.192,2.162)--(4.214,2.162)%
  --(4.236,2.162)--(4.258,2.162)--(4.280,2.162)--(4.302,2.162)--(4.324,2.162)--(4.346,2.162)%
  --(4.368,2.172)--(4.390,2.172)--(4.412,2.172)--(4.434,2.172)--(4.456,2.174)--(4.478,2.174)%
  --(4.500,2.174)--(4.521,2.174)--(4.543,2.174)--(4.565,2.174)--(4.587,2.174)--(4.609,2.174)%
  --(4.631,2.174)--(4.653,2.174)--(4.675,2.181)--(4.697,2.181)--(4.719,2.181)--(4.741,2.181)%
  --(4.763,2.181)--(4.785,2.181)--(4.807,2.181)--(4.828,2.181)--(4.850,2.183)--(4.872,2.183)%
  --(4.894,2.193)--(4.916,2.193)--(4.938,2.193)--(4.960,2.193)--(4.982,2.193)--(5.004,2.193)%
  --(5.026,2.193)--(5.048,2.193)--(5.070,2.193)--(5.092,2.193)--(5.114,2.195)--(5.135,2.195)%
  --(5.157,2.195)--(5.179,2.195)--(5.201,2.195)--(5.223,2.195)--(5.245,2.195)--(5.267,2.195)%
  --(5.289,2.195)--(5.311,2.195)--(5.333,2.207)--(5.355,2.207)--(5.377,2.207)--(5.399,2.207)%
  --(5.421,2.209)--(5.443,2.209)--(5.464,2.228)--(5.486,2.228)--(5.508,2.251)--(5.530,2.251)%
  --(5.552,2.270)--(5.574,2.270)--(5.596,2.272)--(5.618,2.272);
\draw[gp path] (5.662,2.272)--(5.684,2.272)--(5.706,2.284)--(5.728,2.284)--(5.750,2.286)%
  --(5.771,2.286)--(5.793,2.305)--(5.815,2.305)--(5.837,2.307)--(5.859,2.319)--(5.881,2.319)%
  --(5.903,2.337)--(5.925,2.337)--(5.947,2.337)--(5.969,2.337)--(5.991,2.339)--(6.013,2.339)%
  --(6.035,2.339)--(6.057,2.339)--(6.079,2.339)--(6.100,2.339)--(6.122,2.339)--(6.144,2.339)%
  --(6.166,2.339)--(6.188,2.339)--(6.210,2.351)--(6.232,2.351)--(6.254,2.351)--(6.276,2.351)%
  --(6.298,2.358)--(6.320,2.358)--(6.342,2.358)--(6.364,2.358)--(6.386,2.358)--(6.407,2.358)%
  --(6.429,2.358)--(6.451,2.358)--(6.473,2.358)--(6.495,2.358)--(6.517,2.361)--(6.539,2.361)%
  --(6.561,2.361)--(6.583,2.361)--(6.605,2.361)--(6.627,2.361)--(6.649,2.361)--(6.671,2.361)%
  --(6.693,2.370)--(6.714,2.370)--(6.736,2.370)--(6.758,2.370)--(6.780,2.372)--(6.802,2.372)%
  --(6.824,2.372)--(6.846,2.372)--(6.868,2.372)--(6.890,2.372)--(6.912,2.372)--(6.934,2.372)%
  --(6.956,2.372)--(6.978,2.372)--(7.000,2.372)--(7.022,2.372)--(7.043,2.372)--(7.065,2.372)%
  --(7.087,2.375)--(7.109,2.375)--(7.131,2.386)--(7.153,2.386)--(7.175,2.386)--(7.197,2.386)%
  --(7.219,2.394)--(7.241,2.394)--(7.263,2.405)--(7.285,2.405)--(7.307,2.405)--(7.329,2.405)%
  --(7.350,2.408)--(7.372,2.408);
\draw[gp path] (7.570,2.535)--(7.592,2.535)--(7.614,2.535)--(7.636,2.535);
\draw[gp path] (7.679,2.538)--(7.701,2.538)--(7.723,2.538);
\draw[gp path] (7.877,2.549)--(7.899,2.549)--(7.921,2.549)--(7.943,2.549);
\draw[gp path] (8.052,2.571)--(8.074,2.571)--(8.096,2.571)--(8.118,2.571);
\draw[gp path] (8.206,2.585)--(8.228,2.585)--(8.250,2.585);
\gpcolor{color=gp lt color border}
\node[gp node center,font={\fontsize{6.0pt}{7.2pt}\selectfont}] at (5.014,2.421) {$\MI_{2,6,8,10}$};
\gpcolor{rgb color={0.898,0.118,0.063}}
\gpsetlinewidth{1.00}
\gpsetpointsize{4.00}
\gppoint{gp mark 7}{(4.960,2.193)}
\gpcolor{color=gp lt color border}
\node[gp node center,font={\fontsize{6.0pt}{7.2pt}\selectfont}] at (8.304,2.813) {$\MI_{3,5,6,8,10}$};
\gpcolor{rgb color={0.898,0.118,0.063}}
\gppoint{gp mark 7}{(8.250,2.585)}
\gpcolor{color=gp lt color border}
\node[gp node center,font={\fontsize{6.0pt}{7.2pt}\selectfont}] at (8.304,2.813) {$\MI_{3,5,6,8,10}$};
\gpcolor{rgb color={0.898,0.118,0.063}}
\gppoint{gp mark 7}{(8.250,2.585)}
\gpcolor{color=gp lt color border}
\draw[gp path] (0.552,2.750)--(0.552,1.375)--(8.447,1.375)--(8.447,2.750)--cycle;
\gpdefrectangularnode{gp plot 3}{\pgfpoint{0.552cm}{1.375cm}}{\pgfpoint{8.447cm}{2.750cm}}
\end{tikzpicture}

%% file: fig/fig_views.tex
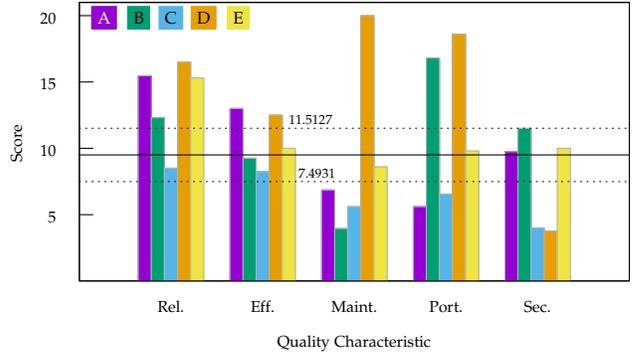
\begin{figure}
 	\input{viewCxP}
\caption{Quality Scores by Characteristic\label{fig:viewCxP}}
\end{figure}

%% file: plot/viewCxP.tex
\begin{tikzpicture}[gnuplot]
\path (0.000,0.000) rectangle (9.000,5.000);
\gpcolor{color=gp lt color border}
\gpsetlinetype{gp lt border}
\gpsetdashtype{gp dt solid}
\gpsetlinewidth{1.00}
\draw[gp path] (1.136,1.867)--(1.316,1.867);
\node[gp node right,font={\fontsize{6.0pt}{7.2pt}\selectfont}] at (0.952,1.867) {$5$};
\draw[gp path] (1.136,2.750)--(1.316,2.750);
\node[gp node right,font={\fontsize{6.0pt}{7.2pt}\selectfont}] at (0.952,2.750) {$10$};
\draw[gp path] (1.136,3.632)--(1.316,3.632);
\node[gp node right,font={\fontsize{6.0pt}{7.2pt}\selectfont}] at (0.952,3.632) {$15$};
\draw[gp path] (1.136,4.515)--(1.316,4.515);
\node[gp node right,font={\fontsize{6.0pt}{7.2pt}\selectfont}] at (0.952,4.515) {$20$};
\node[gp node center,font={\fontsize{6.0pt}{7.2pt}\selectfont}] at (2.355,0.677) {{\Reliability}};
\node[gp node center,font={\fontsize{6.0pt}{7.2pt}\selectfont}] at (3.573,0.677) {{\Efficiency}};
\node[gp node center,font={\fontsize{6.0pt}{7.2pt}\selectfont}] at (4.792,0.677) {{\Maintainability}};
\node[gp node center,font={\fontsize{6.0pt}{7.2pt}\selectfont}] at (6.010,0.677) {{\Portability}};
\node[gp node center,font={\fontsize{6.0pt}{7.2pt}\selectfont}] at (7.229,0.677) {{\Security}};
\draw[gp path] (1.136,4.691)--(1.136,0.985)--(8.447,0.985)--(8.447,4.691)--cycle;
\node[gp node left,font={\fontsize{5.0pt}{6.0pt}\selectfont}] at (3.799,3.171) {11.5127};
\node[gp node left,font={\fontsize{5.0pt}{6.0pt}\selectfont}] at (3.928,2.461) {7.4931};
\node[gp node center,rotate=-270,font={\fontsize{6.0pt}{7.2pt}\selectfont}] at (0.292,2.838) {Score};
\node[gp node center,font={\fontsize{6.0pt}{7.2pt}\selectfont}] at (4.791,0.215) {Quality Characteristic};
\node[gp node right,font={\fontsize{8.0pt}{9.6pt}\selectfont}] at (1.761,4.510) {{\ProdA}};
\gpfill{rgb color={0.580,0.000,0.827}} (1.919,0.985)--(2.094,0.985)--(2.094,3.716)--(1.919,3.716)--cycle;
\gpcolor{color=gp lt color axes}
\draw[gp path] (1.919,0.985)--(1.919,3.715)--(2.093,3.715)--(2.093,0.985)--cycle;
\gpfill{rgb color={0.580,0.000,0.827}} (3.138,0.985)--(3.313,0.985)--(3.313,3.280)--(3.138,3.280)--cycle;
\draw[gp path] (3.138,0.985)--(3.138,3.279)--(3.312,3.279)--(3.312,0.985)--cycle;
\gpfill{rgb color={0.580,0.000,0.827}} (4.356,0.985)--(4.531,0.985)--(4.531,2.196)--(4.356,2.196)--cycle;
\draw[gp path] (4.356,0.985)--(4.356,2.195)--(4.530,2.195)--(4.530,0.985)--cycle;
\gpfill{rgb color={0.580,0.000,0.827}} (5.575,0.985)--(5.750,0.985)--(5.750,1.974)--(5.575,1.974)--cycle;
\draw[gp path] (5.575,0.985)--(5.575,1.973)--(5.749,1.973)--(5.749,0.985)--cycle;
\gpfill{rgb color={0.580,0.000,0.827}} (6.793,0.985)--(6.968,0.985)--(6.968,2.707)--(6.793,2.707)--cycle;
\draw[gp path] (6.793,0.985)--(6.793,2.706)--(6.967,2.706)--(6.967,0.985)--cycle;
\gpcolor{color=gp lt color border}
\node[gp node right,font={\fontsize{8.0pt}{9.6pt}\selectfont}] at (2.202,4.510) {{\ProdB}};
\gpfill{rgb color={0.000,0.620,0.451}} (2.093,0.985)--(2.268,0.985)--(2.268,3.157)--(2.093,3.157)--cycle;
\gpcolor{color=gp lt color axes}
\draw[gp path] (2.093,0.985)--(2.093,3.156)--(2.267,3.156)--(2.267,0.985)--cycle;
\gpfill{rgb color={0.000,0.620,0.451}} (3.312,0.985)--(3.487,0.985)--(3.487,2.618)--(3.312,2.618)--cycle;
\draw[gp path] (3.312,0.985)--(3.312,2.617)--(3.486,2.617)--(3.486,0.985)--cycle;
\gpfill{rgb color={0.000,0.620,0.451}} (4.530,0.985)--(4.705,0.985)--(4.705,1.682)--(4.530,1.682)--cycle;
\draw[gp path] (4.530,0.985)--(4.530,1.681)--(4.704,1.681)--(4.704,0.985)--cycle;
\gpfill{rgb color={0.000,0.620,0.451}} (5.749,0.985)--(5.924,0.985)--(5.924,3.951)--(5.749,3.951)--cycle;
\draw[gp path] (5.749,0.985)--(5.749,3.950)--(5.923,3.950)--(5.923,0.985)--cycle;
\gpfill{rgb color={0.000,0.620,0.451}} (6.967,0.985)--(7.142,0.985)--(7.142,3.015)--(6.967,3.015)--cycle;
\draw[gp path] (6.967,0.985)--(6.967,3.014)--(7.141,3.014)--(7.141,0.985)--cycle;
\gpcolor{color=gp lt color border}
\node[gp node right,font={\fontsize{8.0pt}{9.6pt}\selectfont}] at (2.643,4.510) {{\ProdC}};
\gpfill{rgb color={0.337,0.706,0.914}} (2.267,0.985)--(2.443,0.985)--(2.443,2.486)--(2.267,2.486)--cycle;
\gpcolor{color=gp lt color axes}
\draw[gp path] (2.267,0.985)--(2.267,2.485)--(2.442,2.485)--(2.442,0.985)--cycle;
\gpfill{rgb color={0.337,0.706,0.914}} (3.486,0.985)--(3.661,0.985)--(3.661,2.442)--(3.486,2.442)--cycle;
\draw[gp path] (3.486,0.985)--(3.486,2.441)--(3.660,2.441)--(3.660,0.985)--cycle;
\gpfill{rgb color={0.337,0.706,0.914}} (4.704,0.985)--(4.880,0.985)--(4.880,1.974)--(4.704,1.974)--cycle;
\draw[gp path] (4.704,0.985)--(4.704,1.973)--(4.879,1.973)--(4.879,0.985)--cycle;
\gpfill{rgb color={0.337,0.706,0.914}} (5.923,0.985)--(6.098,0.985)--(6.098,2.142)--(5.923,2.142)--cycle;
\draw[gp path] (5.923,0.985)--(5.923,2.141)--(6.097,2.141)--(6.097,0.985)--cycle;
\gpfill{rgb color={0.337,0.706,0.914}} (7.141,0.985)--(7.317,0.985)--(7.317,1.692)--(7.141,1.692)--cycle;
\draw[gp path] (7.141,0.985)--(7.141,1.691)--(7.316,1.691)--(7.316,0.985)--cycle;
\gpcolor{color=gp lt color border}
\node[gp node right,font={\fontsize{8.0pt}{9.6pt}\selectfont}] at (3.084,4.510) {{\ProdD}};
\gpfill{rgb color={0.902,0.624,0.000}} (2.442,0.985)--(2.617,0.985)--(2.617,3.897)--(2.442,3.897)--cycle;
\gpcolor{color=gp lt color axes}
\draw[gp path] (2.442,0.985)--(2.442,3.896)--(2.616,3.896)--(2.616,0.985)--cycle;
\gpfill{rgb color={0.902,0.624,0.000}} (3.660,0.985)--(3.835,0.985)--(3.835,3.192)--(3.660,3.192)--cycle;
\draw[gp path] (3.660,0.985)--(3.660,3.191)--(3.834,3.191)--(3.834,0.985)--cycle;
\gpfill{rgb color={0.902,0.624,0.000}} (4.879,0.985)--(5.054,0.985)--(5.054,4.516)--(4.879,4.516)--cycle;
\draw[gp path] (4.879,0.985)--(4.879,4.515)--(5.053,4.515)--(5.053,0.985)--cycle;
\gpfill{rgb color={0.902,0.624,0.000}} (6.097,0.985)--(6.272,0.985)--(6.272,4.268)--(6.097,4.268)--cycle;
\draw[gp path] (6.097,0.985)--(6.097,4.267)--(6.271,4.267)--(6.271,0.985)--cycle;
\gpfill{rgb color={0.902,0.624,0.000}} (7.316,0.985)--(7.491,0.985)--(7.491,1.648)--(7.316,1.648)--cycle;
\draw[gp path] (7.316,0.985)--(7.316,1.647)--(7.490,1.647)--(7.490,0.985)--cycle;
\gpcolor{color=gp lt color border}
\node[gp node right,font={\fontsize{8.0pt}{9.6pt}\selectfont}] at (3.525,4.510) {{\ProdE}};
\gpfill{rgb color={0.941,0.894,0.259}} (2.616,0.985)--(2.791,0.985)--(2.791,3.687)--(2.616,3.687)--cycle;
\gpcolor{color=gp lt color axes}
\draw[gp path] (2.616,0.985)--(2.616,3.686)--(2.790,3.686)--(2.790,0.985)--cycle;
\gpfill{rgb color={0.941,0.894,0.259}} (3.834,0.985)--(4.009,0.985)--(4.009,2.751)--(3.834,2.751)--cycle;
\draw[gp path] (3.834,0.985)--(3.834,2.750)--(4.008,2.750)--(4.008,0.985)--cycle;
\gpfill{rgb color={0.941,0.894,0.259}} (5.053,0.985)--(5.228,0.985)--(5.228,2.504)--(5.053,2.504)--cycle;
\draw[gp path] (5.053,0.985)--(5.053,2.503)--(5.227,2.503)--(5.227,0.985)--cycle;
\gpfill{rgb color={0.941,0.894,0.259}} (6.271,0.985)--(6.446,0.985)--(6.446,2.715)--(6.271,2.715)--cycle;
\draw[gp path] (6.271,0.985)--(6.271,2.714)--(6.445,2.714)--(6.445,0.985)--cycle;
\gpfill{rgb color={0.941,0.894,0.259}} (7.490,0.985)--(7.665,0.985)--(7.665,2.751)--(7.490,2.751)--cycle;
\draw[gp path] (7.490,0.985)--(7.490,2.750)--(7.664,2.750)--(7.664,0.985)--cycle;
\gpcolor{color=gp lt color border}
\draw[gp path] (1.136,4.691)--(1.136,0.985)--(8.447,0.985)--(8.447,4.691)--cycle;
\gpcolor{rgb color={0.000,0.000,0.000}}
\gpsetlinetype{gp lt axes}
\gpsetdashtype{gp dt axes}
\draw[gp path](1.136,2.307)--(8.447,2.307);
\gpsetdashtype{gp dt solid}
\draw[gp path](1.136,2.662)--(8.447,2.662);
\gpsetdashtype{gp dt axes}
\draw[gp path](1.136,3.017)--(8.447,3.017);
\gpdefrectangularnode{gp plot 1}{\pgfpoint{1.136cm}{0.985cm}}{\pgfpoint{8.447cm}{4.691cm}}
\end{tikzpicture}

%% file: tex/discussion.tex
This case study is a preliminary indication of the
validity of {\OAFIR}.  Obviously, more experiments with
{\OAFIR} on different \SPL{s} are needed to
increase the method's validity.  Still, the \SPL{} on which {\OAFIR} was evaluated is a standard
\SPL{}.  The method did not rely on any special
attributes, such as a specific domain.  Therefore,
one can expect {\OAFIR} to work on other \SPL{s}.


Quality scores~(\Fref{viewCxP}) were nicely distributed, with some
features having repeatedly low scores, some features
with repeatedly high scores, and some features
having varying scores. This made the selection
step (\Sref{STEP:VI}) challenging (no
trivial optimal architecture), demonstrating the strength
of {\OAFIR}.

Selecting just five systems from the full \SPL{}
succeeded in identifying quality gaps at the \SPL{}
level in a timely manner, without needing to evaluate
each and every product \cite{elorza2008evaluacion}.
However, after matching the results with other
products in the \SPL{}, another recurring
quality gap was identified.  Therefore, finding the
optimal number of products to analyze in a \SPL{}
should be further researched.

The \SPL{} management accepted the
recommendations from our case study, putting effort
and budget to implement the suggested modifications.
This is further indication of the validity of {\OAFIR}
and relevance to industry.

%% file: tex/related.tex

\subsection{Single Product Evaluation Methods}

\emph{Domain-specific software architecture comparison
model} (DoSAM) \cite{bergner2005dosam} is an \AOM{}
that uses a metric comparison~(\TraitVI), in multiple
quality features~(\TraitVII), and is cost-aware~(\TraitVIII).  Weights of
each quality feature are calculated mathematically.
Overall, this method enables a systematic evaluation
of a single product.

However, DoSAM depends on the alternative
architectures to be given as input.  If the optimal
architecture is not in the input alternatives, it
will not be found.  Also, it does not easily scale to
the full list of quality features, as it demands a
specific analysis of the impact of each architectural
element on each of the quality features.

\subsection{Product Line Evaluation Methods}
%
%
Several related works extend their \AOM{} to target \PLA{}
	evaluation~\cite{etxeberria2005product},
	including:
    \emph{holistic product line architecture assessment}~(HoPLAA)~\cite{olumofin2007holistic},
    \emph{distributed scenario-based architecture analysis method}~(D-SAAM)~\cite{Graaf:2005:EES},
    \emph{extended architecture tradeoff analysis method}~(EATAM)~\cite{Kim:2008:EAA},
and
    \emph{calidad del producto y del proceso software}~(CaLiPro)~\cite{elorza2008evaluacion}.

These \AOM{s} apply a single product evaluation
	method, similar to that of SAAM~\cite{Kazman:1994:SMA}
	and ATAM~\cite{Kazman:2000:AMA}, but in two phases:
	first to the \SPL{}'s core architecture,
	and then to each of the different products.
	They identify suitability and quality gaps of
	the whole \SPL{} (\TraitIX) with multiple
	quality features (\TraitVII). 
However, they have
	to be expanded with quantitative metric techniques
	(\TraitVI) in order to be used in the \AOP{} of product
	line architectures~\cite{olumofin2007holistic},
	Moreover, these methods do not support generation of
	alternative architectures, which is a task far from trivial
	in a product line.

%
%
\emph{Quality-driven product architecture derivation and
	improvement}~(QuaDAI)~\cite{Gonzalez:2013:DVM} is an \AOM{}
	for a \SPL{}.
The \SPL{} architect identifies
	the existing product architecture and its relations with the
	quality requirements, and measures adherence to those requirements
	using external metric methods.
The \SPL{} architect has to come up with a set
	of possible modifications.  Then, a domain expert ranks each
	possible modification on a scale of 1 to 9.
Iteratively,
	each modification is applied to the product architecture in
	an effort to meeting all the quality requirements.

While QuaDAI is designed for a \SPL{}, its steps
	are focused on a single product.
However, it
does not consider trade-offs
	between single product modifications and shared assets or changes
	to the core architecture.
Therefore, it
	cannot find an optimal \PLA{}, but rather a
	local optimization for each product.
%
%


%% file: tex/conclusion.tex

In this paper we introduced {\OAFIR}, an end-to-end
method for the {\AOP}.  Instead of using incompatible
\AOM{s} for each step, {\OAFIR} takes the architect
from the initial requirements all the way to the
optimal architecture.


Using {\OAFIR}, the architect analyzes multiple
quality characteristics, selected from ISO/IEC~25010.
The method is 
fitted to the \SPL{} by defining features and assigning
weights.  {\OAFIR} also supports the identification
of cross-\SPL{} weaknesses, the generation
of respective cross-\SPL{} solutions, and
the reuse of one product's effective solution across
the \SPL{}.
The significance of the various quality
characteristics to the organization, as declared by
the stakeholders, is kept separate from the effect
of the various feature, as identified by technical
means, while both influence the selection of the
optimal architecture.

We presented a case study where {\OAFIR} was applied
to a production \SPL{} in an industrial setting.
Weaknesses of the \SPL{} and its specific
systems were identified, a list of potential
modifications was generated, and an optimal \PLA{} was produced. 

Once the {\AOP} is formulated with the discretion
of stakeholders being an input to the process rather
than part of it, it might be possible to automate
the process.  {\OAFIR} thus lays the foundation for
an industry-level automatic generation of software
architectures and product line architectures---a
topic left for future work.